\documentclass[12pt]{iopart}

\usepackage{epsfig}
\usepackage{graphicx}% Include figure files

\begin{document}

\newcommand{\lesssim}   {\mathrel{\mathop{\kern 0pt \rlap
  {\raise.2ex\hbox{$<$}}}
  \lower.9ex\hbox{\kern-.190em $\sim$}}}
\newcommand{\gsim}   {\mathrel{\mathop{\kern 0pt \rlap
  {\raise.2ex\hbox{$>$}}}
  \lower.9ex\hbox{\kern-.190em $\sim$}}}

\newcommand{\be}{\begin{equation}}
\newcommand{\ee}{\end{equation}}
\newcommand{\ba}{\begin{eqnarray}}
\newcommand{\ea}{\end{eqnarray}}
\def\bone{$B^{(1)}$}
\def\bone{B^{(1)}}
\def\etal{{\it et al.~}}
\def\eg{{\it e.g.~}}
\def\ie{{\it i.e.~}}
\def\DM{dark matter~}
\def\DE{dark energy~} 
\def\GC{Galactic center~} 
\def\susy{SUSY~}

\title{GZK Photons Above 10 EeV}

\author{Graciela Gelmini$^a$, Oleg Kalashev$^{a,b}$ and Dmitry
  V. Semikoz$^{b,c}$ }

\address{$^a$ Department of Physics and Astronomy, UCLA, Los Angeles,
CA 90095-1547, USA}
\address{$^b$INR RAS, 60th October 
Anniversary pr. 7a, 117312 Moscow, Russia}
\address{$^c$ APC, College de France, 11 pl. 
Marcelin Berthelot, Paris 75005, France }

%\vspace{0.5truecm}
\begin{abstract}
We calculate the flux of ``GZK-photons", namely the flux of 
photons produced by  extragalactic nucleons through the resonant
photoproduction of pions,  the so called GZK effect.  This flux  depends on the
UHECR spectrum on Earth, of the spectrum of nucleons
emitted at the sources, which we characterize by its slope
 and maximum energy, on the distribution of sources and on  the intervening cosmological backgrounds,
 in particular the magnetic field and radio backgrounds.  For the first time  we calculate the GZK photons produced by nuclei. We calculate the possible range of the GZK photon fraction of the total UHECR flux 
 for the AGASA   and the HiRes spectra. We find that for nucleons produced at the sources it could be as large as a  few \%  and as low as 10$^{-4}$ above 10$^{19}$~eV. For nuclei produced at the sources the maximum photon fraction is a factor of 2 to 3 times smaller above 10$^{19}$~eV but the minimum could be much smaller than for nucleons.  We also comment on cosmogenic neutrino fluxes.
\end{abstract}

%\hfill UCLA/07/TEP/?? }
\maketitle

%\vspace{1truecm}

\section{Introduction}

The cosmic rays with energies beyond the Greisen-Zatsepin-Kuzmin (GZK)
cutoff~\cite{gzk} at $4\times 10^{19}$~eV present a challenging outstanding
puzzle in astroparticle physics and 
cosmology~\cite{agasa, agasa_spec, hires, hires_mono_spec}. Nucleons
cannot be significantly deflected by the magnetic fields
  of our  galaxy for energies above the ``ankle", i.e.  above
10$^{18.5}$~eV. This and the absence of a correlation of arrival directions
with the galactic plane indicate that, if nucleons are the primary particles
of the ultra high energy cosmic rays (UHECR), these nucleons should be of
extragalactic origin. However, nucleons  as well as photons with energies above 
$5 \times 10^{19}$~eV could not reach Earth from a distance beyond 50 to 100
Mpc~\cite{50Mpc, 40Mpc} and no sources have been so far found within this distance.
 Nucleons scatter off the cosmic microwave background
(CMB) photons with a resonant photoproduction of pions $p\gamma \rightarrow
\Delta^* \rightarrow N\pi$, where the pion carries away $\sim 20\%$ of the
original nucleon energy.  The mean free path for this reaction is only
$6$~Mpc. Photons with comparable energy pair-produce electrons and positrons
on the radio background and the uncertainty in this background translates 
into uncertainty in the photon energy-attenuation length.

Intervening sheets of large scale intense extra galactic magnetic fields
(EGMF), with intensities $B \sim 0.1 -1\times 10^{-6}$~G, could provide
sufficient angular deflection for protons to explain the lack of observed
sources in the directions of arrival of UHECR. However, recent realistic
simulations of the expected large scale EGMF, show that strong deflections
could only occur when particles cross galaxy clusters. Except in the regions
close to the Virgo, Perseus and Coma clusters the magnetic fields are
not larger than 3$\times 10^{-11}$~G~\cite{dolag2004} and the deflections
expected are not important (however  see Ref.~\cite{Sigl:2004yk}).

Whether particles can be emitted with the necessary energies by astrophysical
accelerators, such as active galactic nuclei, jets or extended lobes of radio galaxies, 
or even extended object such as colliding galaxies and
clusters of galaxies, is still an open question. The size and possible
magnetic and electric fields of these astrophysical sites make it plausible
for them to produce UHECR at most up to energies of $10^{21}$~eV. Larger emission
energies would require a reconsideration of possible acceleration models or
sites.

Heavy nuclei are an interesting possibility for UHECR primaries, since they
could be produced at the sources with larger maximum energies (proportional to
their charges) and would more easily be deflected by intervening magnetic
fields. On the other hand, both AGASA and HiRes data favor 
a dominance of light hadrons, consistent with
being all protons, in the composition of UHECR above 10$^{19}$~eV. 
These data are consistent with models in
which  all UHECR above $10^{18}$ eV are due to 
extragalactic protons~\cite{berezinsky2002}.

A galactic component of the UHECR flux, which could be important 
up to energies  10$^{19}$~eV, should consist of heavy nuclei, given the
 lack of
correlation with the galactic plane of events at this energy (outside the
galactic plane galactic protons would be deflected by a maximum of 15-20$^o$
at this energies~\cite{galactic_magn_field}).

The GZK cutoff at $4\times 10^{19}$~eV seems not to be present in the data of
the AGASA ground array~\cite{agasa} but it appears in the data of the HiRes
air fluorescence detector~\cite{hires, hires_mono_spec}.  
This controversy ca be addressed by the Pierre Auger
Observatory~\cite{Auger}, a hybrid combination of charged particles detectors
and fluorescence telescopes, as it continues to accumulate data.

The GZK process produces pions. From the decay of $\pi^{\pm}$ one obtains
neutrinos.  These  ``cosmogenic neutrinos"  (sometimes called these days
``GZK neutrinos") have been extensively studied, from
1969~\cite{bere} onward 
(see for example~\cite{reviewGZKneutrinos,Semikoz:2003wv} and
references therein), and constitute one of the main high energy signals
expected in neutrino telescopes, such as ICECUBE~\cite{ICECUBE} 
ANITA~\cite{ANITA} and  SALSA~\cite{SALSA} 
or space based observatories such as EUSO~\cite{EUSO} and
OWL~\cite{OWL}.  From the decay of $\pi^0$ we obtain photons, 
``GZK photons", each with about 0.1 of
the original proton energy, which have been known to be a subdominant
component of the UHECR since the work of Wdowczyk {\it et al.} in the early
1970's~\cite{wdowczyk}.
In 1990  it was suggested that if the  extragalactic radio background
and magnetic fields are small ($B< 3 \times 10^{-11}$ G)
GZK photons could dominate over protons and explain the
super-GZK events~\cite{Aharonian1990}. The dependence
of the GZK photon flux on extragalactic magnetic fields was later
studied in Ref.~\cite{SiglOlinto95}. The argument of Ref.~\cite{Aharonian1990}
 and its dependence on extragalactic magnetic fields
was again discussed~\cite{astro_photons} in connection with the possible
correlation of UHECR arrival directions with BL Lacertae objects~\cite{Tinyakov:2001nr}.
However, to our knowledge, the first complete study of the expected
fluxes of  GZK photons, including their dependence on
the initial proton fluxes, distribution of proton sources and UHECR spectrum,
besides intervening backgrounds, was done in Ref.~\cite{gzk_photon}, and completed
here with an improved statistical analysis (first  used in Ref.~\cite{Gelmini:2007sf}).

With the advent of the Pierre Auger Observatory, we expect to have in the near future
the high statistic data that may allow to study a subdominant component of
UHECR consisting of photons~\cite{bertoux, Risse:2007sd}.  Auger has already set bounds on the photon fraction above 10$^{19}$~eV~\cite{Abraham:2006ar} and better bounds are expected soon.
The GZK photons provide a complementary handle to
GZK neutrinos and other signatures to try to determine the spectrum and
composition of the UHECR. The flux of GZK photons is necessarily correlated
with the flux of  cosmogenic
 neutrinos, although the former is affected by the
radio background and EGMF values which do not affect the latter.
Auger will hopefully see photons or place a limit 
on the photon fraction  at the level of  a few \% or below. If complemented by an extended
 northern array~\cite{Auger_North}, a sensitivity level of below 0.1\% could be reached within a few years of full operation~\cite{Risse:2007sd}.
Here we  would like to address the physical implications of such detection
or limits.

In this paper we fit the assumed UHECR spectrum 
above 4 $\times 10^{19}$eV solely with primary nucleons and the 
GZK photons they produce. The GZK photon flux
  depends on the UHECR spectrum assumed,  the slope and maximum energy of
the primary nucleon flux, the distribution of sources and the
 intervening backgrounds~\cite{gzk_photon}.
We take a phenomenological approach in choosing the range of the several relevant
parameters which determine the GZK photon flux, namely we take for each 
of them a range of values mentioned in the literature, 
without attempting to assign them to particular sources or acceleration mechanisms.
  We also study here for the first time the GZK photon
  flux produced if nuclei are emitted at the sources. 
  In this case we
    fit the AGASA or HiRes spectrum  above
4 $\times 10^{19}$eV with the nuclei and nucleons resulting from the disintegration
of the primary nuclei and the GZK photons they produce.
  To compare them with the ratios produced by pure protons,  we present the GZK photon
  ratios  in the simplified case in which either only He, or O
   or Fe would be produced at the sources.
   
    The ankle in the UHECR spectrum at energies $10^{18}$eV - $10^{19}$ eV 
can be explained either by $e^\pm$ production by  extragalactic protons or by a
change from one component of the UHECR spectrum to another.
The latter explanation assumes the existence of a low energy component (LEC) 
when necessary to fit the
assumed UHECR at energies below $10^{19}$~eV.   This LEC 
can be dominated by galactic Fe or by a different population of lower
energy extragalactic nucleons. This last possibility can  still  be consistent
 with the proton-dominated composition observed by HiRes. Here we do not address
the issue of what the  LEC is.  We only assume that, if it exists,  it becomes negligible
at energies above  4 $\times 10^{19}$eV, the energy above which we fit the
 data. In addition we impose that the spectrum we predict is never above the
  measured spectrum at energies below 4 $\times 10^{19}$eV. 
The LEC in any event does not contribute to the flux of GZK photons since
 it is important at energies under the
threshold for photo-pion production.

In order to find the expected range of the GZK photon flux, we fit  
either the AGASA or the HiRes data  above  4 $\times 10^{19}$eV either
 minimizing or maximizing the number of GZK protons produced.  We find
  (see Figs.~\ref{E-plots} and \ref{E-plots-zmin}) that  (assuming exclusively
   protons are emitted at the sources) the GZK photon fraction of the total 
   integrated  UHECR flux could reach a few \% above 10$^{19}$~eV  and  10\%
   above 10$^{20}$~eV,
   or be between one (for AGASA) and several (for HiRes) orders of magnitude smaller,
 under the level that could be detected at Auger South alone. In fact, we find (as in  Ref.~\cite{Gelmini:2007sf} and Ref.~\cite{Sigl:2007ea})
that the  photon fraction in cosmic rays at
energies above $10^{19}$ eV could be as low as $O(10^{-4})$. Photon fluxes so small could only be detected in
future experiments like Auger North plus South~\cite{Auger_North, Risse:2007sd}, EUSO~\cite{EUSO} and OWL~\cite{OWL}.
  In Fig.~\ref{E-plots} a zero minimum distance to the sources is assumed. i.e. a minumum distance much smaller than all relevant interaction lenghts. In Fig.~\ref{E-plots-zmin} a minimum distance  to the sources of 50 Mpc (actually a minimum redshift 0.01) is assumed. 
As clearly shown in Figs.~\ref{Emax-plots} (where $z_{\rm min}=0$) and
Fig.~\ref{Emax-plots-zmin} (where $z_{\rm min}$  is allowed to vary between 0 and  0.01) the GZK photon fractions
depend strongly on the maximum energy of the protons initially emitted at the sources. 
Just for comparison in  Figs.~\ref{Emax-plots}  and ~\ref{E-plots}  we also show  the range of GZK-photon 
fractions expected if purely He or purely Fe nuclei  (also O in Figs.~\ref{Emax-plots}) were emitted at the sources, 
 and in these cases the maximum GZK photon fractions expected are smaller.

 The detection of GZK photons would open the way for  UHECR photon astronomy.
The detection of a larger photon flux than expected for GZK photons given the
particular UHECR spectrum,
would imply the emission of photons at the
source or new physics. New physics is involved in Top-Down models,
produced as an alternative to acceleration models to explain the
 origin of the highest energy cosmic rays. All  Top-Down models
 predict photon dominance at the highest energies. If photons are not seen,
 Auger will  place  interesting bounds
on  production models.

The plan of the paper is the following. In Section II,  we explain how
 we model the sources and the propagation of particles.
In Section III, we calculate  the maximum and 
minimum GZK photon fractions expected either with the AGASA spectrum 
or with the HiRes spectrum.
 In Section IV  we show  that very different  cosmogenic neutrino fluxes
  could be associated with UHECR spectra with either  minimum and maximum GZK photon fractions.

\section{Modeling of the sources and particle  propagation }

We use a numerical code described in Ref.~\cite{kks1999}  
to compute the flux of GZK
photons produced by an homogeneous distribution of sources emitting originally
only protons or nuclei.   This is the
same numerical code as in Ref.~\cite{gzk_photon}, with a few
modifications.  

The code uses the  kinematic equation approach and
calculates the propagation of nuclei, nucleons, stable leptons  and photons
 using the standard dominant processes, explained for example in
 Ref.~\cite{reviews1}).
For nucleons, it takes into account single and multiple pion production and $e^{\pm}$ pair production on
 the CMB, infrared/optical and radio  backgrounds, neutron
 $\beta$-decays and the expansion of the Universe. For nuclei, it takes into account
pion  production,  $e^{\pm}$ pair production and photodissociation  through
scattering with  infra-red and CMB photons.  For photons, the
 code includes  $e^{\pm}$ pair production, $\gamma + \gamma_B
 \rightarrow e^+ e^-$ and double  $e^{\pm}$ pair production $\gamma +
 \gamma_B \rightarrow e^+ e^-  e^+ e^- $,
 processes. For electrons and positrons, it takes into account inverse Compton
 scattering, $e^\pm + \gamma_B \rightarrow e^\pm \gamma$,
triple pair production,  $e^\pm + \gamma_B \rightarrow
 e^\pm e^+  e^- $ ,  and synchrotron energy loss on extra galactic
 magnetic fields (EGMF).  All these reactions are  discussed in detail
 for example in the Ph.D. thesis of  S. Lee \cite{propagLeeSigl} and  that of
 O. Kalashev~\cite{kks1999}. The propagation of  nucleons and
the  electron-photon cascades are calculated self-consistently, namely
 secondary (and higher generation) particles arising in all reactions
 are  propagated alongside the primaries. The  hadronic
 interactions of nucleons are now derived from the well established SOPHIA event
generator~\cite{Mucke:1999yb}, more accurate in the multi-pion
 regime  than the old code in Ref.~\cite{kks1999}.  For the photodisintegration coefficients of nuclei 
we use the approximation first introduced in Ref.~\cite{puget} and then revised
in Ref.~\cite{Stecker:1998ib}. As a check,  we reproduce the energy loss length of iron
obtained in Ref.~\cite{Stecker:1998ib} when using the same
 infrared-optical spectrum  used in  Ref.~\cite{Stecker:1998ib}. The simulation of the
electron-photon cascade
development was verified by detailed comparisons of its results with those
obtained with an analogous code developed by an  independent
group~\cite{propagLeeSigl}. It has already been used in a series of papers
dedicated to cosmic rays and astrophysics~\cite{zburst_problem,
 Neronov:2002se, Kalashev:2002kx, Semikoz:2003wv}.
 
 UHE particles lose their energy in interactions with the electro-magnetic
background, which consists of CMB,  radio, infra-red and optical (IRO) components,
as well as EGMF.  Protons are sensitive essentially to the CMB only, while for
UHE photons and nuclei the radio and IRO components are respectively
important,  besides the CMB. Notice that the radio background is not yet
well known and that our conclusions 
depend strongly on the background assumed. We
include three models for the radio background: the  background based on
estimates by Clark {\it et al.}~\cite{clark} and the two models of
Protheroe and Biermann~\cite{PB}, both predicting a larger background than the
first.   For the IRO background component we used the model
of Ref.~\cite{Stecker:2005qs}.
 The infra-red and optical background 
is not important for the production of GZK photons from primary nucleons  at
 high energies  and  their absorption.
This background is important to  transport the energy of secondary photons
 in the cascade process 
from the  0.1 - 100 TeV energy range  to the  0.1-100 GeV energy range 
observed by EGRET,  and 
the resulting flux in this energy range is not sensitive to details
of the IRO background models.  The IRO background is also important
for the photodisintegration of nuclei,  thus affects  the photon fluxes predicted by models
with sources emitting nuclei.

It is believed that the magnetic fields in  clusters
 can be generated from a primordial ``seed'' if it has a
comoving magnitude $B \sim 10^{-12}$~G \cite{Dolag:2002,dolag2004}.  The
evolution of the EGMF together with the large scale structure of the Universe has
been simulated recently by two groups using independent numerical procedures
\cite{Sigl:2004yk,dolag2004}. Magnetic field strengths significantly larger
than 10$^{-10}$~G were found only within large clusters of galaxies.  In our
simulations we vary the magnetic field strength in the range $B = 10^{-11} -
10^{- 9}$~G, assuming an 
unstructured field along the propagation path.

Notice that  if neutrons  are produced at the sources, the results at
 high energies are  identical to those obtained with protons. 
 The interactions of neutrons and protons with the intervening
backgrounds are identical and when  a neutron decays practically all
 of its energy goes to the final  proton (while the electron and neutrino are
produced with energies 10$^{17}$~eV or lower).

The resulting GZK photon flux depends on several
astrophysical parameters.  These parametrize the initial proton flux, the
distribution of sources, the radio background and the EGMF.
As it is usual, we take the spectrum of an individual UHECR source  to
be of  the form:
\be  F(E) = f E^{-\alpha} ~~\Theta (E_{\rm max} -E)~,
 \label{proton_flux}
\ee
where $f$ provides the flux normalization, $\alpha$ is the spectral  index and
$E_{\rm max}$ is the maximum energy to which protons can be
accelerated at the source.

 We assume a standard cosmological model with  a Hubble constant
$H=70$~km~s$^{-1}$~Mpc$^{-1}$, a dark energy density (in units of the
critical density) $\Omega_{\Lambda}= 0.7$ and a dark matter density
$\Omega_{\rm m}=0.3$. The total source density in this model can be
defined by
\be n(z) = n_0 (1+z)^{3+m}~  \Theta (z_{\max}-z) \Theta (z-z_{\min}) \,,
\label{sources}
\ee
where $m$  parameterizes the source density evolution, in such a   way
that $m=0$ corresponds to non-evolving sources with constant density
per comoving volume,  and $z_{\min}$ and $z_{\max}$ are respectively
the redshifts of the closest and most distant sources.  We have fixed
 $m=0$ in this paper (except in Fig.~\ref{cosmogenic}) because the
 high energy photons come from close by, thus the effect of the evolution of sources is small
(we estimate that this may introduce an uncertainty of the order of 10\% in the photon fluxes we find).
 Sources with $z>2$ have a negligible contribution  to the UHECR flux
above $10^{18}$~eV.

We are implicitly assuming that the sources are astrophysical, since these are the
 only ones which could produce solely protons (or neutrons) and nuclei as UHECR
 primaries. Astrophysical acceleration mechanisms often result in $\alpha
 \gsim 2$~\cite{AS2}, however, harder spectra, $\alpha \lesssim 1.5$ are also
 possible, see e.g. Ref.~\cite{AS1.5}. In reality, the spectrum may differ from a power-law,
 it may even have  a peak at high energies~\cite{peaks}. 
 AGN cores could  accelerate protons with induced
electric fields, similarly to what happens in a
linear accelerator, and this mechanism would produce  an 
almost monoenergetic proton flux, with energies as high as $10^{20}$~eV or
higher~\cite{mono}. Here, we consider the
 power law index to be in the  range  $1 \le \alpha \le 2.7$.
 An injected
proton spectrum with $\alpha \geq 2.5$ does not require an extra contribution
to fit the UHECR data, except at very low energies 
$E<10^{18}$ eV~\cite{Berezinsky:2002vt}.
 For $\alpha \le 2$ an extra low energy component
(LEC) is required to fit the UHECR data at $E<10^{19}$~eV.
 The flux of super-GZK protons 
  (and thus the flux of GZK photons too) depends strongly on the power law
index $\alpha$ of the initial injected  proton flux: it is lower for larger values
 of $\alpha$.  The dependence of the GZK photon flux on  the maximum 
 energy $E_{\rm max}$ is more significant as $\alpha$ decreases. Here we will consider values of  $E_{\rm max}$ up to $10^{21}$~eV.
 
 Most of the energy in GZK photons cascades down to below the pair production 
threshold for photons on the CMB and infrared backgrounds.  In general, for $\alpha<2$ 
the diffuse extragalactic gamma-ray flux measured by EGRET~\cite{EGRET}
 at GeV energies may impose a constraint on the GZK photon flux at high energies, 
 which we  take into account and found not relevant for any of the models we study here.

The value of $z_{\min}$ is connected to the
density of sources. Quite often in the literature the minimal distance to 
the sources is assumed to
be negligible (i.e. comparable to the interaction length).  We  also consider 
non-zero minimun distances of up to 50 Mpc  (actually $z_{\rm
min}= 0.01$), as inferred from the
small-scale clustering of events seen in the AGASA data~\cite{AGASA_clusters}. 
Contrary to AGASA, HiRes does not see a clustering component in its own
data~\cite{HiRes_clusters}. The combined dataset shows that clustering still
exists, but it is not as significant as in the data of AGASA
alone~\cite{agasa_hires}.  Note, that the non-observation of clustering in the
HiRes stereo data does not contradict the result of AGASA, because of the
small number of events in the sample~\cite{agasa_hires_ok}.
Assuming proton primaries and a small EGMF (following Ref.~\cite{dolag2004}),
it is possible to infer the density of the 
sources~\cite{sources, agasa_hires_ok} 
from the clustering component of UHECR. AGASA data alone
suggest a source density of $2\times 10^{-5}$~Mpc$^{-3}$, which makes
plausible the existence of one source within 25 Mpc of us. However, the HiRes
negative result on clustering requires a larger density of sources and, as a
result, a smaller distance to the nearest one of them. Larger 
values of the EGMF
(as found in Ref.~\cite{Sigl:2004yk}), and/or some fraction of iron
in the UHECR, have the effect of reducing the required number of sources and,
consequently, increasing the expected distance to the nearest one.

\section{Expected range of GZK photon fractions}

In this section we estimate the maximum and minimum GZK photon
flux expected if the UHECR spectrum is that of AGASA~\cite{agasa_spec} or that 
 HiRes~\cite{hires_mono_spec}.  We proceed using the method explained in Ref.~\cite{Gelmini:2007sf}.

As shown in Ref.~\cite{gzk_photon}, the largest GZK photon fractions in UHECR
 happen for small values of $\alpha$, large values of $E_{\rm max}$,
 and small intervening backgrounds. The smallest
GZK photon fluxes happen with the opposite choices. 
To take into account the effect of the intervening backgrounds, here we  fit the UHECR data assuming
either a maximal intervening  background (the largest radio background of Protheroe
and Biermann~\cite{PB} and large EGMF, $B=10^{-9}G$) or a minimal 
intervening  background (the radio background of Clark {\it et al.}~\cite{clark}
and small EGMF, $B=10^{-11}G$), with many different injected spectra. 
 We assume the injected spectrum in  Eq.~\ref{proton_flux},
a uniform distribution of sources with a density as in Eq.~\ref{sources} with
$z_{\rm max} =3$ and,  to start with, $z_{\rm min} =0$ and $m=0$.
We consider then many different spectra resulting from changing   the slope
$\alpha$ and the maximum energy $E_{\max}$ in 
Eq.~\ref{proton_flux} within the ranges 
$1 \leq \alpha\leq 2.9$ and $10^{20} {\rm eV}\leq E_{\max}\leq  1.3 \times 10^{21}$~eV  
in steps $\alpha_n=1+0.1 n$, with $n=1$ to 19 and 
$E_{\ell}=Z \times 10^{19}eV \times 2^\ell$, with $\ell=0$ to 10, where $Z$ is the electric charge
of the particle injected ($Z=1$ for protons, but later we apply the same procedure to nuclei as well).
 For each one of the models so obtained we compute the  predicted UHECR spectrum
by summing up the contributions of protons plus GZK photons
 arriving to us from all sources.

In order to compare  the predicted flux with the data, we take also into account the experimental error in the energy determination
as proposed in Ref.~\cite{Albuquerque:2005nm}. We take a
lognormal distribution for the error in the  energy reconstructed by the
 experiment with respect to the true value
of energy of the UHECR coming into the atmosphere. To find the expected flux we convolute the spectrum predicted by each model with the  lognormal distribution in energy with
the width given by the HiRes energy error $\Delta E /E = 17$ \% \cite{hires_mono_spec} and the AGASA
energy error $\Delta E /E = 25$ \%~\cite{agasa_spec} (the parameter $\sigma$ in Eq. (5) of 
Ref.~\cite{Albuquerque:2005nm}, the standard deviation of log$_{10} E$, is
$\sigma = (\Delta E / E)/$ ln(10)$ \simeq (\Delta E / E)/ 2.3$).
This procedure results in small but not negligible changes in the
predicted spectra which are then compared to the  observed spectrum. In particular, there are 
events predicted with energy larger than the maximum energy $E_{\max}$. Somewhat arbitrarily
we consider the energy beyond which no event is predicted to be $(1 + 10 \Delta E/E)E_{\max}$.
Moreover, in the case of the AGASA spectrum, we take into account that there is a 1.2 factor between the energy of a photon event and the energy measured if the event is reconstructed assuming it is a proton~\cite{1.2factorAGASA, AgasaYakutskLimit}. Thus we divide the energy of the predicted  GZK photon energy by 1.2 before comparing it with the observed AGASA spectrum.
 
  With each predicted spectrum we  fit the UHECR data from 4 $\times 10^{19}$~eV  up to the last published bin  of each spectrum (i.e. the 9
highest energy data  bins of  AGASA  or the 12  highest 
energy   bins of HiRes 1 and 2 combined monocular data) possibly  plus one
 extra bin at  higher energies. This last additional bin with
  zero observed events is added only if  the maximum energy $(1 + 10 \Delta E/E)E_{\max}$
(where $E_{\max}$ is the maximum energy assumed for the  injected spectrum in Eq.~\ref{proton_flux}) is larger than the
maximum energy of the last published bin (i.e. larger than 3.16 $\times10^{20}$~eV for AGASA~\cite{agasa_spec}  and 3.98 $\times10^{20}$~eV for HiRes~\cite{hires_mono_spec}). This additional empty bin extends from the last published experimental point of each observed spectrum (which also are empty) to  $(1 + 10 \Delta E/E)E_{\max}$. We compute
 the expected number of events in this last bin using an exposure  that we derive from the AGASA or HiRes data above 10$^{20}$~eV  and assuming the exposure is 
 energy independent (above 10$^{20}$~eV). 
 This extra bin and the highest energy empty published bins,  take into account   the non-observation of events above the highest occupied energy bin in the data of each
 collaboration, the end-point energy of each spectrum
  (i.e. at $E> 2.3 \times 10^{20}$ eV for AGASA~\cite{agasa_spec} 
and $E> 1.6 \times 10^{20}$~eV for HiRes~\cite{hires_mono_spec}),
although their aperture remains constant with increasing energy. 

\begin{figure}
\includegraphics[width=0.34\textwidth,clip=true,angle=270]{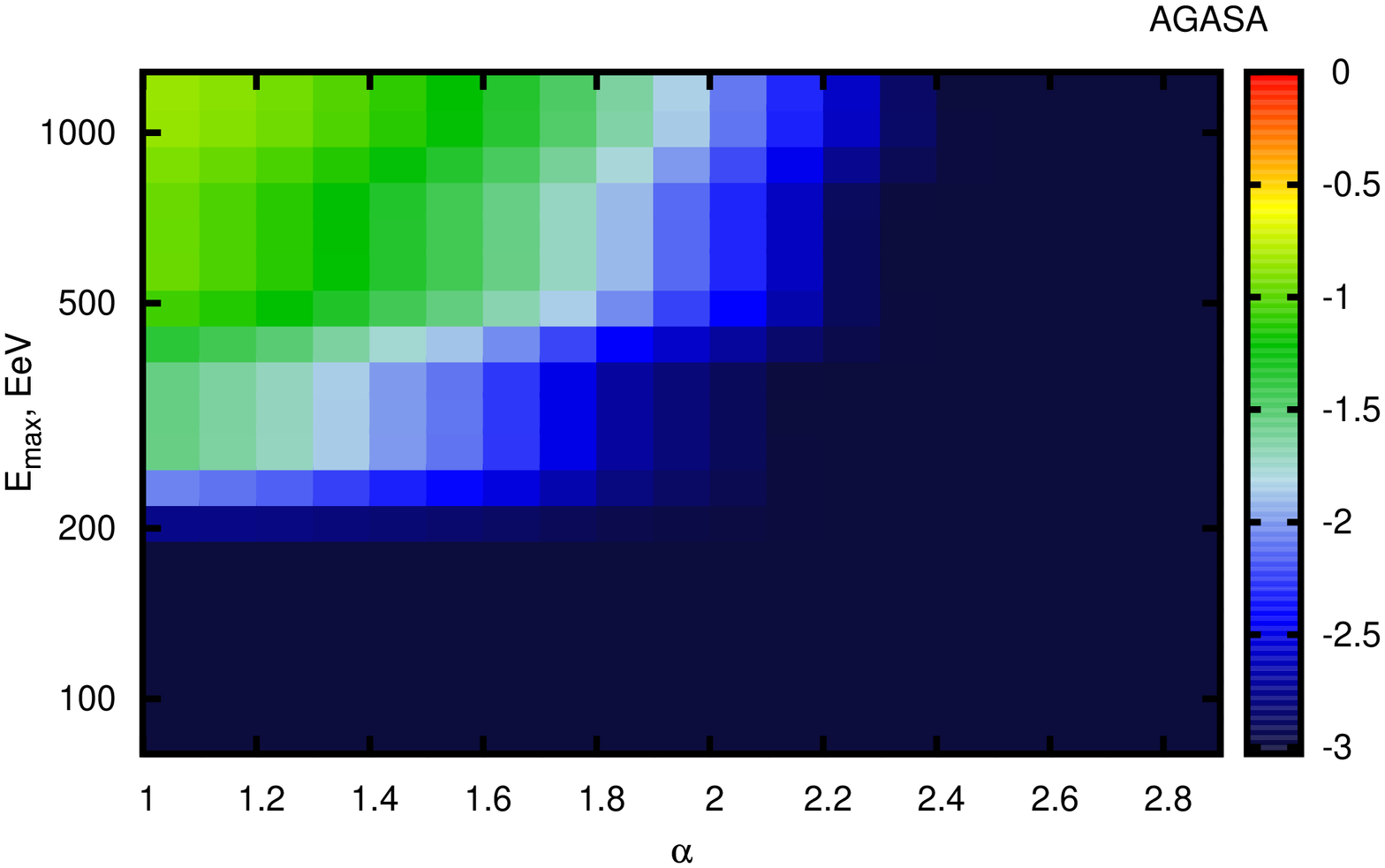}
\includegraphics[width=0.34\textwidth,clip=true,angle=270]{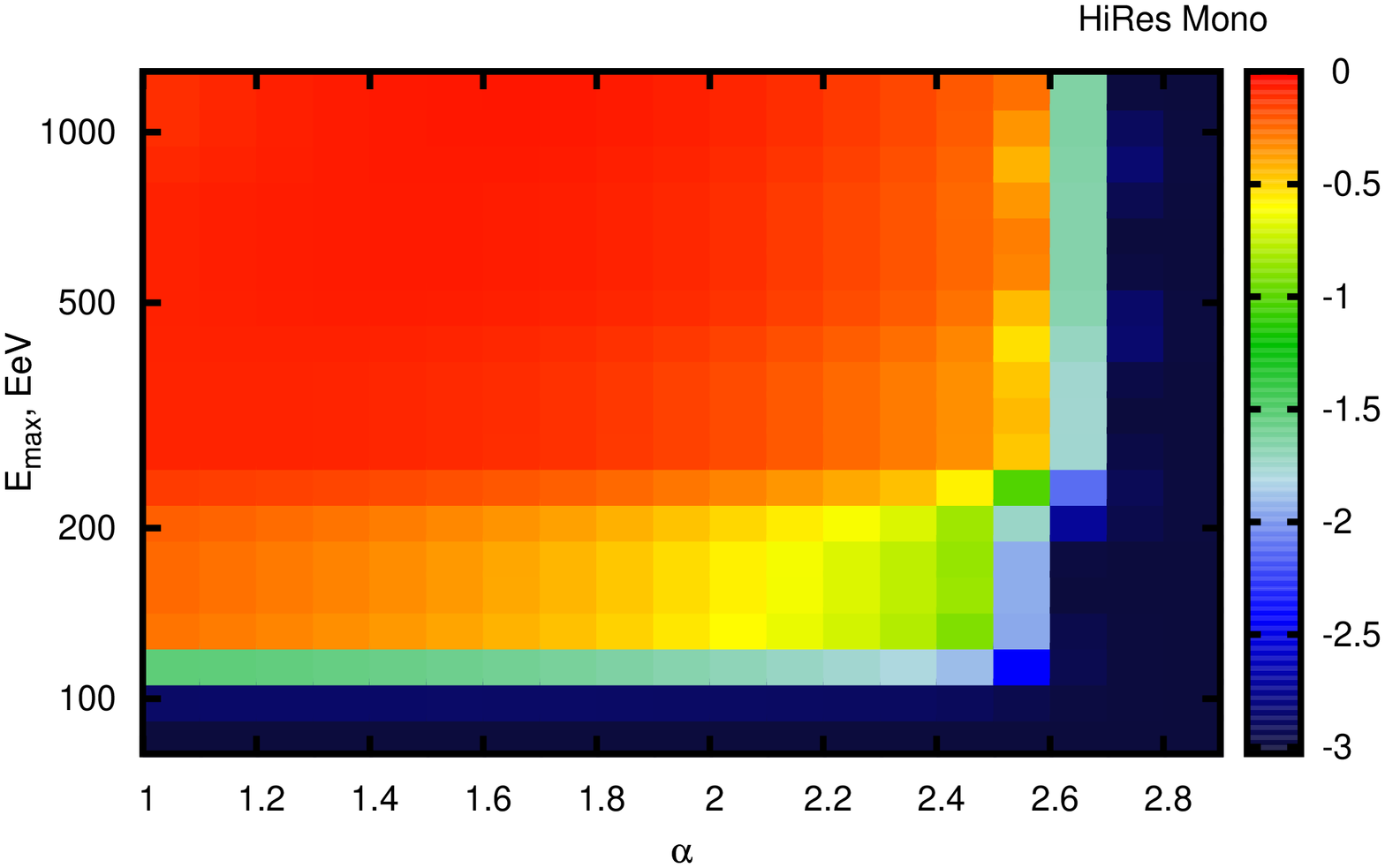}
\caption{
Color coded  $p$-value plots as function of $E_{\max}$ and 
$\alpha$ for the AGASA (left panel) and HiRes (right panel) spectra above 
 4 $\times 10^{19}$~eV. Figures look the same for  maximum
 or  minimum  intervening radio background and EGMF. The  color scale indicates the power index $x$ where $p= 10^{x}$.}
%Fig.1
\label{p-values-filter}
\end{figure}
\begin{figure}[ht]
\includegraphics[width=0.34\textwidth,clip=true,angle=270]{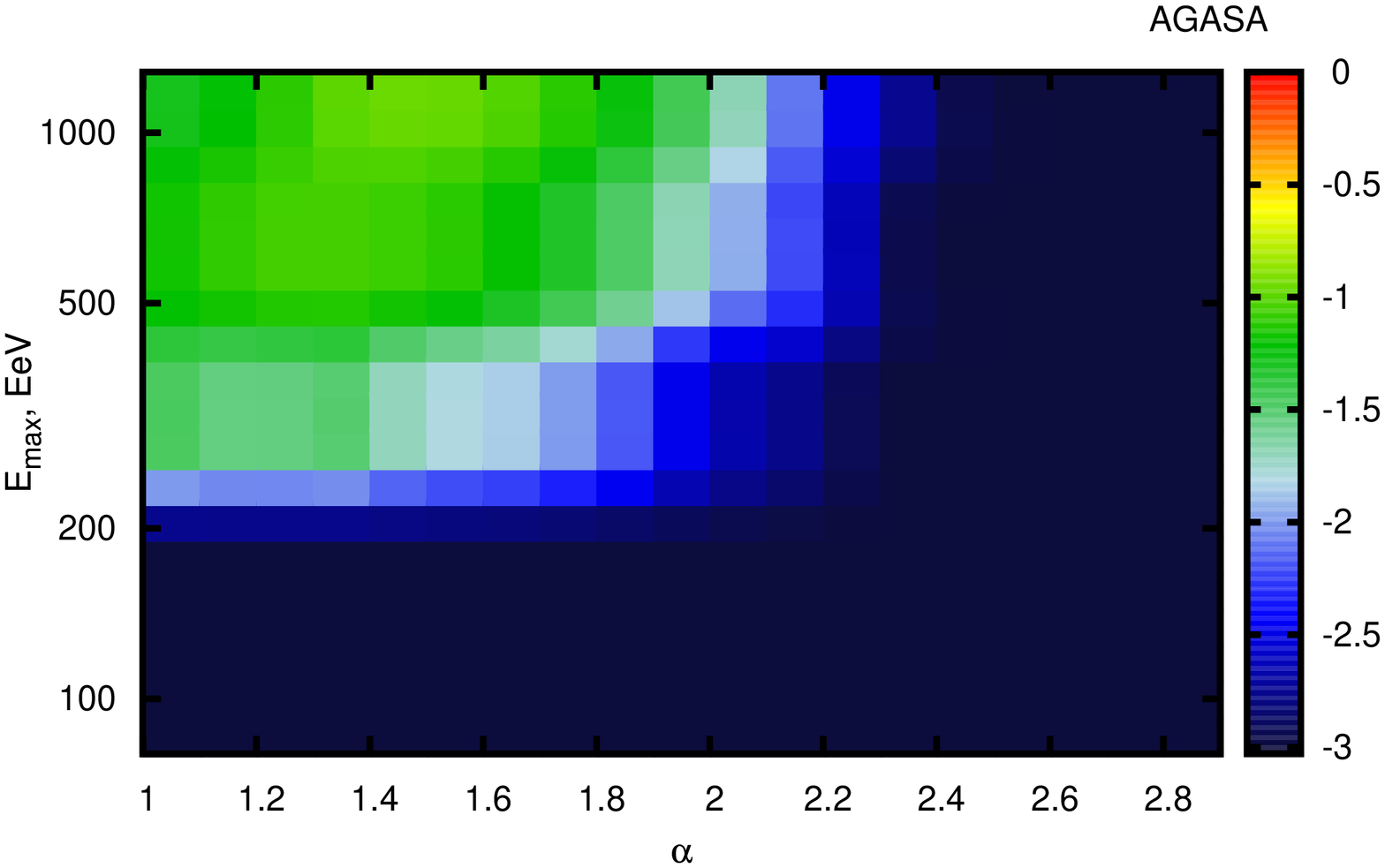}
\includegraphics[width=0.34\textwidth,clip=true,angle=270]{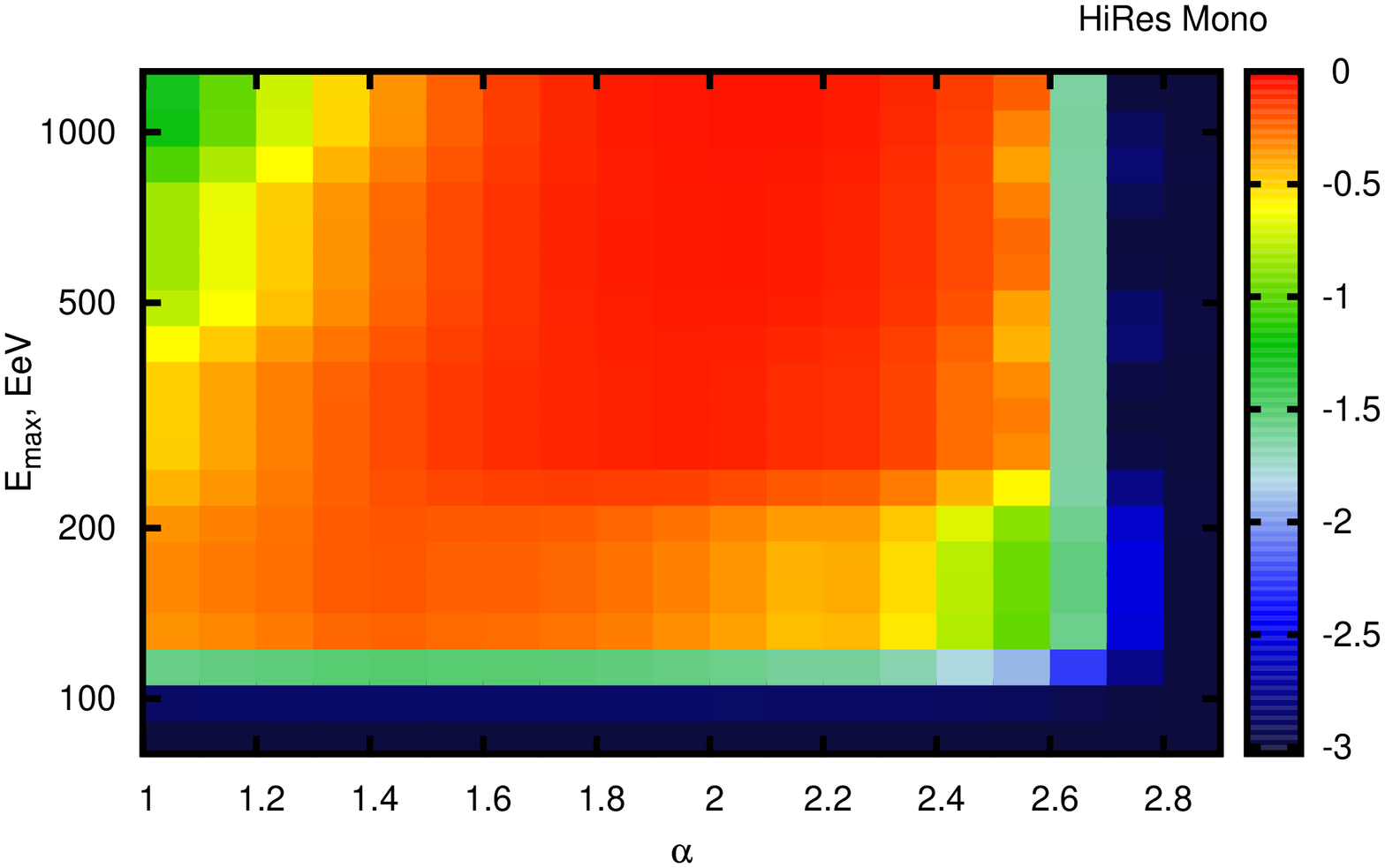}
\caption{Same as Fig.~\ref{p-values-filter} but  maximizing the Poisson likelihood using UHECR data above 2  $\times 10^{19}$~eV  instead of 4  $\times 10^{19}$~eV. Figures look the same for  maximum
 or  minimum  intervening radio background and EGMF}
%Fig. 2
\label{p-values-filter-2}
\end{figure}

To fit the UHECR data with each predicted spectrum we follow a procedure similar to that of Ref.~\cite{Fodor-K-R} applied to the bins just mentioned.  The number of events and exposure data used to produce the latest published HiRes flux figures in Ref.~\cite{hires_mono_spec} are given in Ref.~\cite{HiRes-table}. For
the AGASA spectrum, we reconstruct the measured number of events in each bin from the published data (using the error bars~\cite{Poisson-errors}). In both cases we compare the observed number of events in each bin with the number of events in each bin predicted by each one of the models.
 We choose the value of the
parameter $f$ in Eq.~\ref{proton_flux}, i.e. the amplitude of the injected spectrum, by maximizing the Poisson likelihood function, which is equivalent to minimizing $-2 \ln{\lambda}$, (i.e. the negative of the log likelihood ratio)~\cite{statistics}. This procedure amounts to choosing the value of $f$  so that the mean total number of events predicted (i.e. the sum of the average predicted number of events in all fitted   bins)  is equal to the total number of events observed. We then compute using a Monte Carlo technique the goodness of the fit, or $p$-value of the distribution, defined as the mean fraction of hypothetical experiments (observed spectra)  with the same fixed total number of events which would result in a worse, namely lower, Poisson likelihood than the one obtained (in the maximization procedure that fixed $f$). These hypothetical experiments are chosen at random according to the multinomial distribution of the model (with $f$ fixed as described).  We have checked that this procedure  when applied to bins with large number of events gives the same results as  a Pearson's $\chi^2$ fit, both for the value of the normalization parameter $f$ and for the goodness of fit.  A higher $p$ value corresponds to a better fit, since more hypothetical experimental results would yield a worse fit than the one we obtained.
\begin{figure}
\hspace{4cm}
\includegraphics[width=0.34\textwidth,clip=true,angle=270]{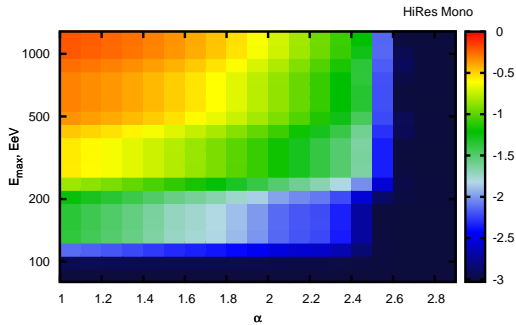}
\caption[...]{
Same as Fig.~\ref{p-values-filter}   but  for $z_{\rm min} = 0.01$
 (insted of $z_{\rm min} = 0$) and only for the HiRes spectrum (there are no acceptable solutions for the AGASA spectrum). Figures look the same for  maximum
 or  minimum  intervening radio background and EGMF}
\label{p-values-filter-zmin} 
%Fig. 3
\end{figure}

We make one additional requirement on the fit that insures that the predicted flux does not exceed  the observed flux at energies below 4 $\times 10^{19}$eV. We use the published fluxes at energies above 3 $\times 10^{18}$~eV for AGASA and 1.8 $\times 10^{17}$~eV for HiRes. For each assumed spectrum (with $f$ fixed  as described above) we calculate the $\chi^2$ for the data at low energies only using the data points in which the predicted flux is above the observed flux (i.e. we take as zero the  contribution to the $\chi^2$ of each data point for which the predicted flux is below the observed flux). We then require the $p$-value of the $\chi^2$ so obtained to be larger than 0.05. This constraint eliminates the lowest values of $\alpha$ and $E_{\rm max}$. The regions eliminated by this requirement are assigned a $p$-value equal to zero in Figs.~\ref{p-values-filter}, \ref{p-values-filter-2} and \ref{p-values-filter-zmin}.

 Fig.~\ref{p-values-filter} shows in a logarithmic scale the color coded $p$-value of the maximum Poisson likelihood  value obtained for each model as function of $E_{\max}$ and  $\alpha$ for the AGASA (left panel) and HiRes (right panel) spectra. The figures look the same if they are produced  with maximum or minimum intervening radio background and EGMF, because the contribution of GZK photons to the predicted spectra is always subdominant.
In the case of AGASA we see that the best fits (those with larger $p$-values) occur at the largest
$\alpha$ and $E_{\rm max}$ we consider, and the fits are never very good: the $p$-value is never larger than 0.2. The best fits to the HiRes data can be better than the best fits to the AGASA data, the $p$-values can reach 0.7,  and lie  in a crescent-moon shaped region at more moderate values of $\alpha$ and $E_{\rm max}$.

  The $p$-values obtained depend on the energy range chosen for the Poisson likelihood fit. In choosing to fit the data at energies 4 $\times 10^{19}$~eV and above without a low energy component (LEC) we are assuming that any LEC necessary to fit the spectrum at lower energies is negligible in this energy range. The goodness of the  fit to the UHECR
data depends  on the lowest energy of the range we choose to fit, because the number of events in the  lower energy bins is large. Had we chosen to fit the UHECR data from 2 $\times 10^{19}$~eV instead we would have obtained the $p$-values which we  show in 
Fig.~\ref{p-values-filter-2}, just for comparison with Fig.~\ref{p-values-filter}. In Fig.~\ref{p-values-filter-2} the region of best models moved to somewhat lower values of   $\alpha$ and $E_{\rm max}$  and their combined largest values (at the top left corner of the figure) are disfavored with respect to Fig.~\ref{p-values-filter}. We believe that a possible LEC may  still be important  to fit the UHECR data at 2 $\times 10^{19}$~eV. For  example, iron would still be strongly deflected by the galactic magnetic fields up  to energies to close to 3   $\times 10^{19}$~eV. Thus, a galactic iron component may be important
up energies just below 4 $\times 10^{19}$~eV.

\begin{figure}
\includegraphics[width=0.34\textwidth,clip=true,angle=270]{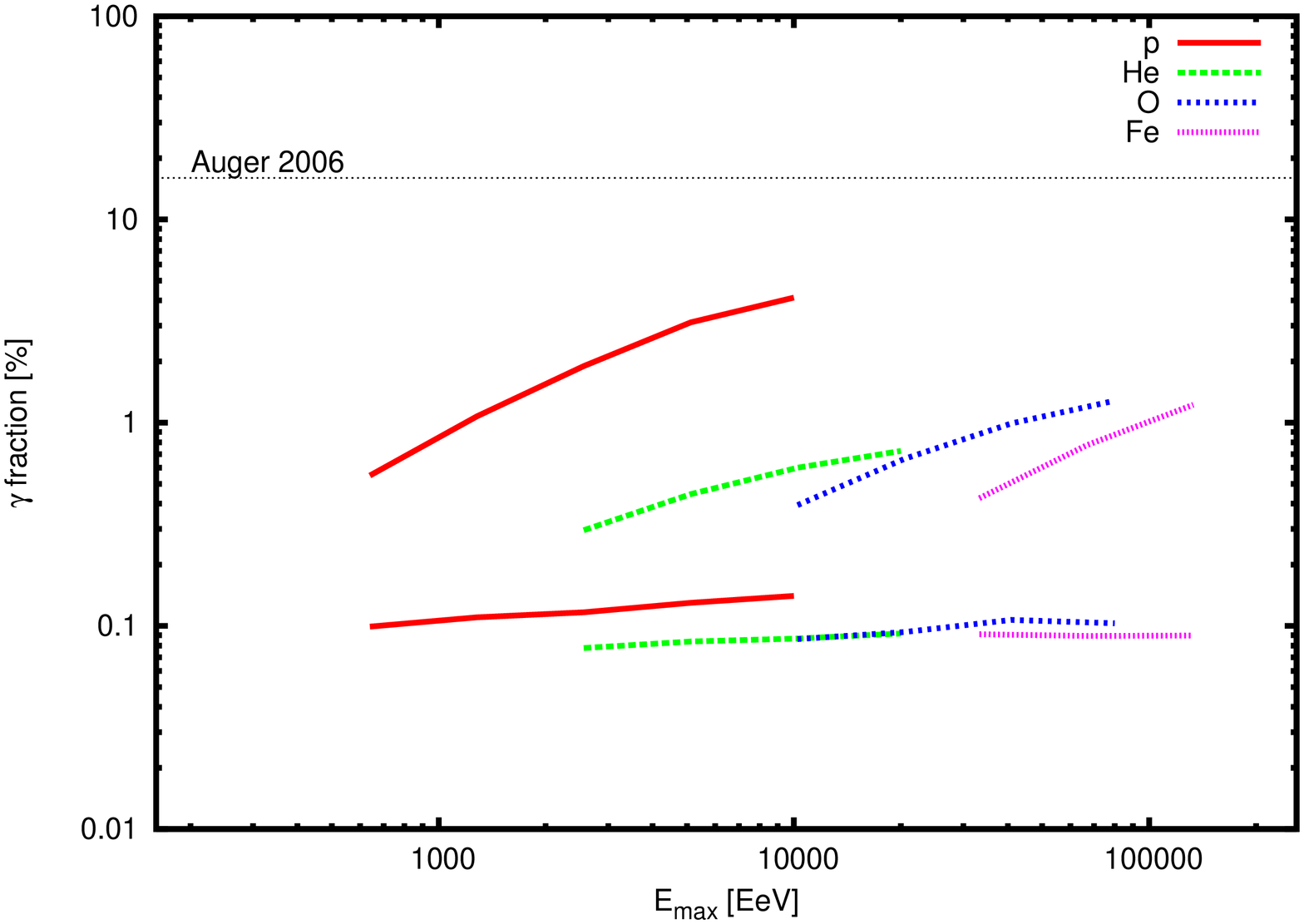}
\includegraphics[width=0.34\textwidth,clip=true,angle=270]{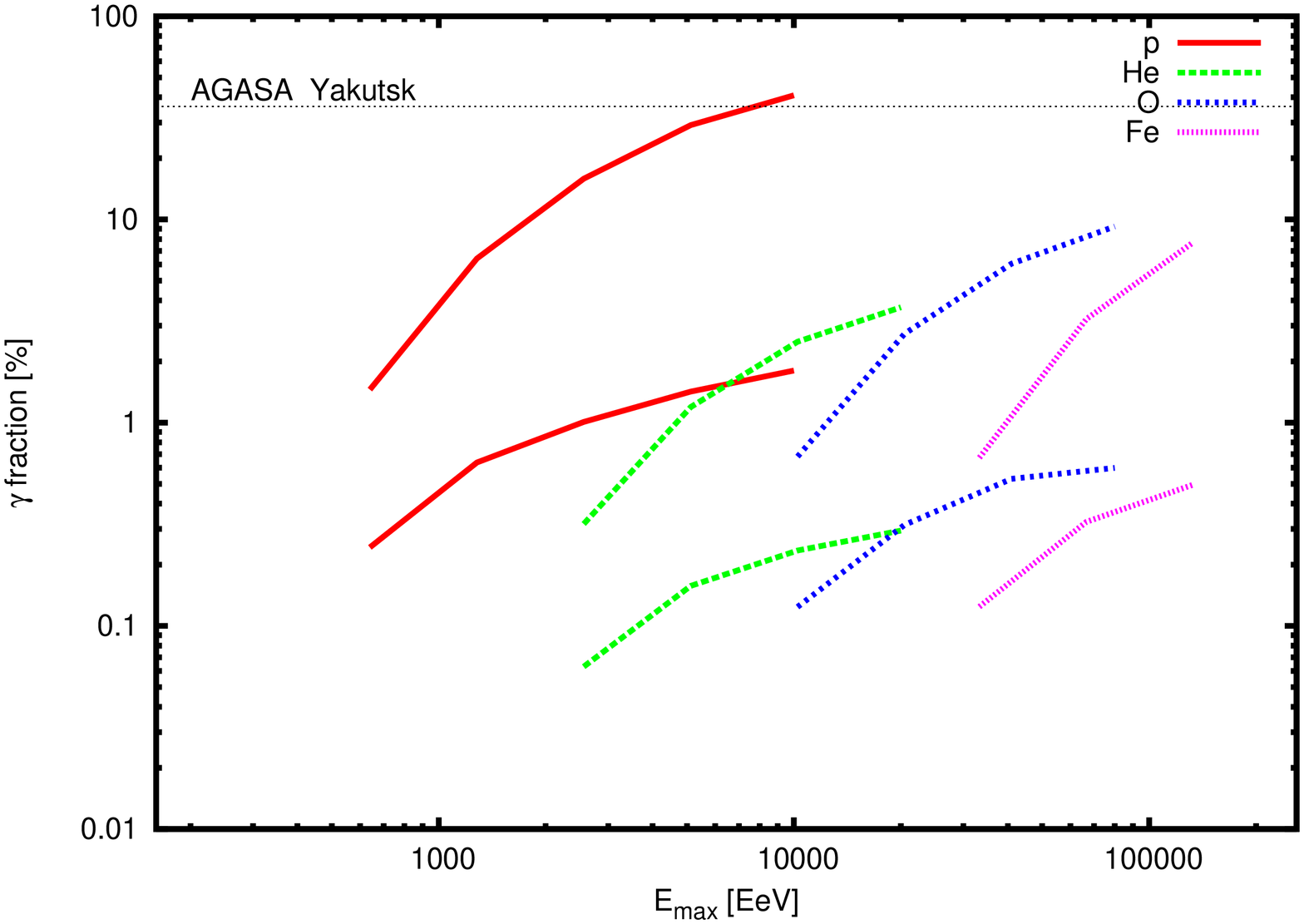}
\includegraphics[width=0.34\textwidth,clip=true,angle=270]{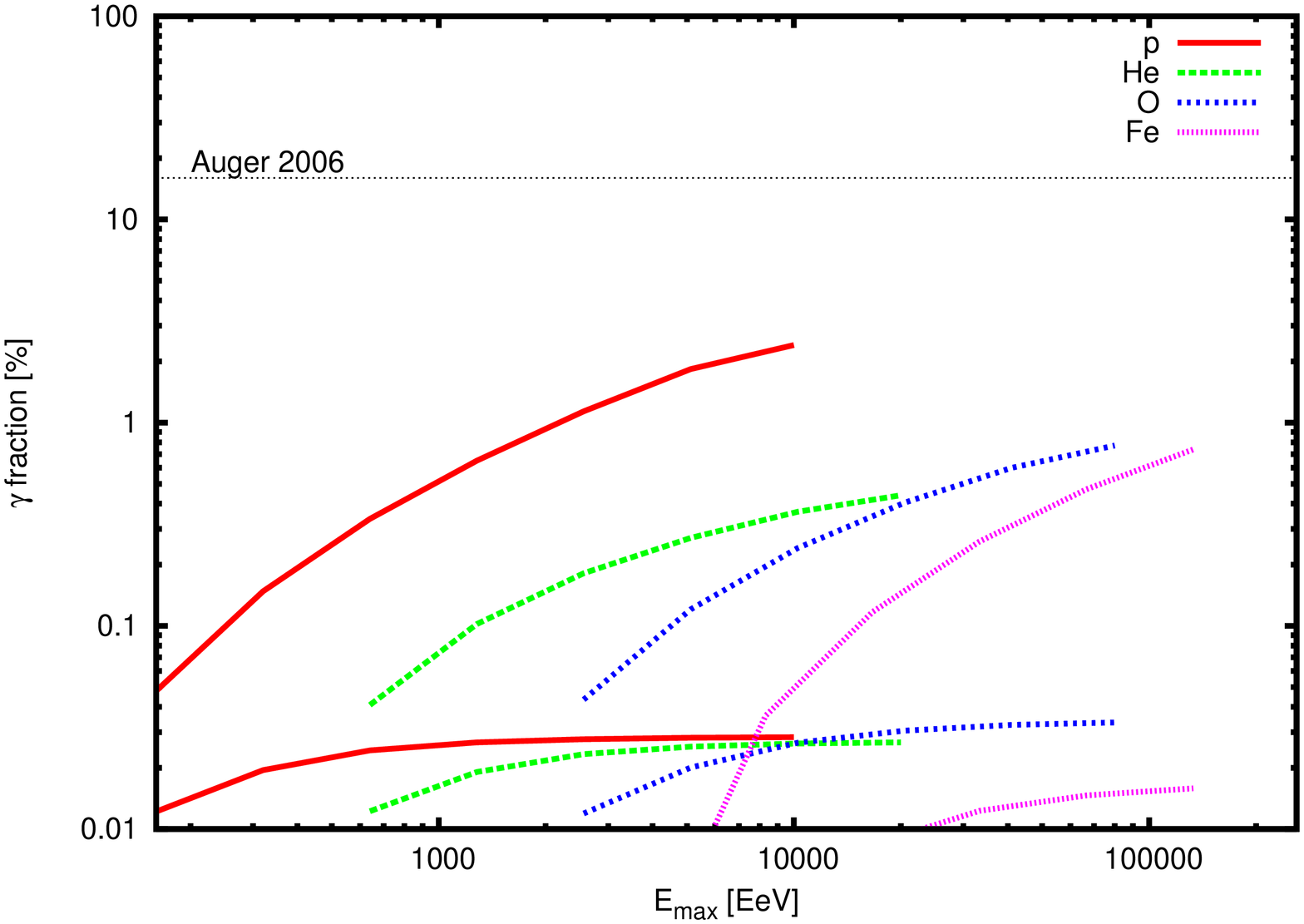}
\includegraphics[width=0.34\textwidth,clip=true,angle=270]{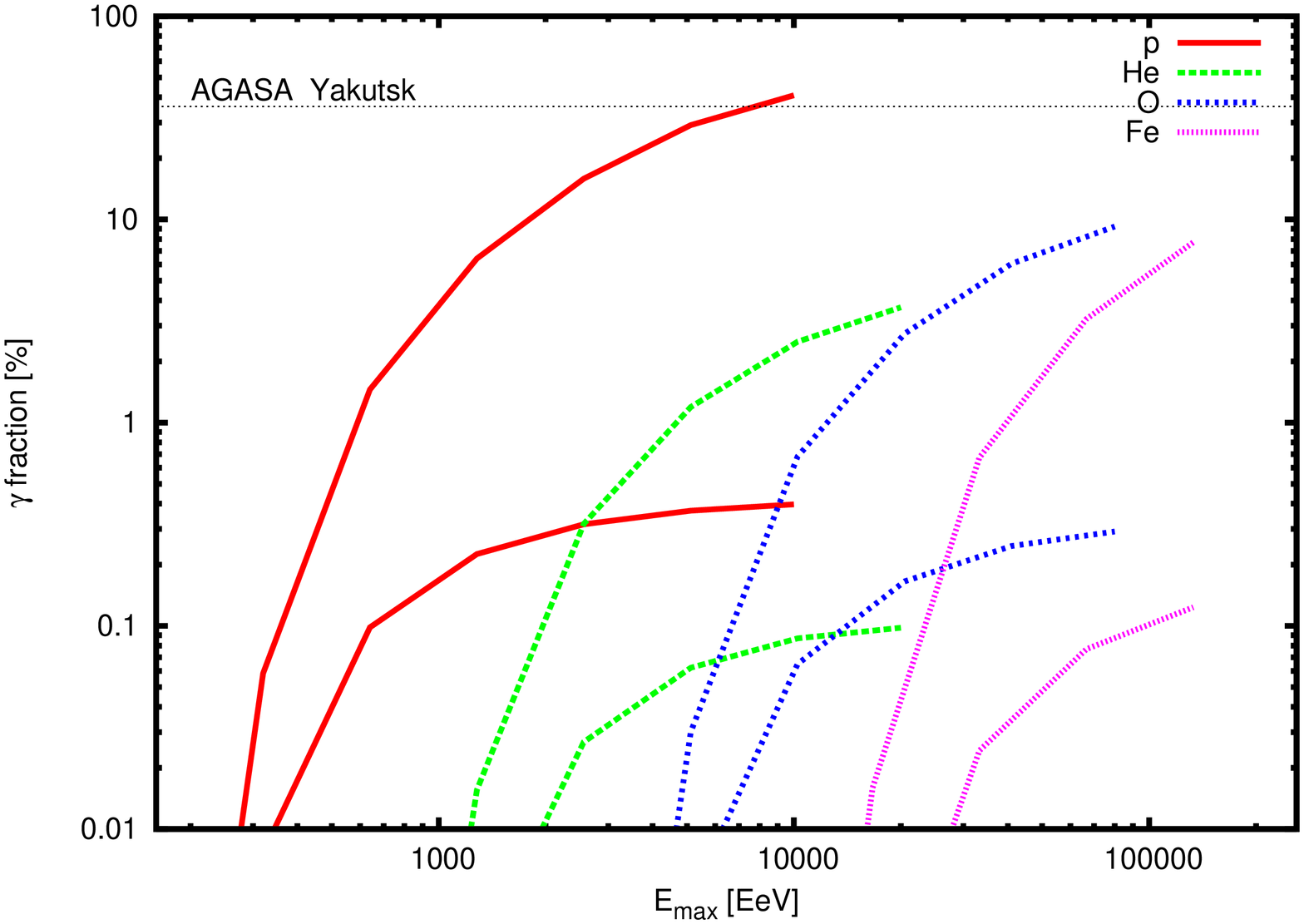}
\caption[...]{Maximum and minimum GZK photon fractions 
as  function of $E_{\rm max}$ between $Z \times 10^{20}$~eV and $Z \times 10^{22}$~eV, given  in percentage of the integrated fluxes above  $E= 1 \times 10^{19}$eV (left panels) and $1 \times 10^{20}$eV (right panels) for AGASA (upper panels) and HiRes (lower panels) respectively, found among the models with $p$-value $>0.05$ in Fig.\ref{p-values-filter}. Here we kept $z_{\rm min} = 0$.
The colors indicate different primaries assumed: red for proton, green for He, blue for O and magenta for Fe.}
\label{Emax-plots} 
%Fig. 4
\end{figure}
\begin{figure}
\includegraphics[width=0.34\textwidth,clip=true,angle=270]{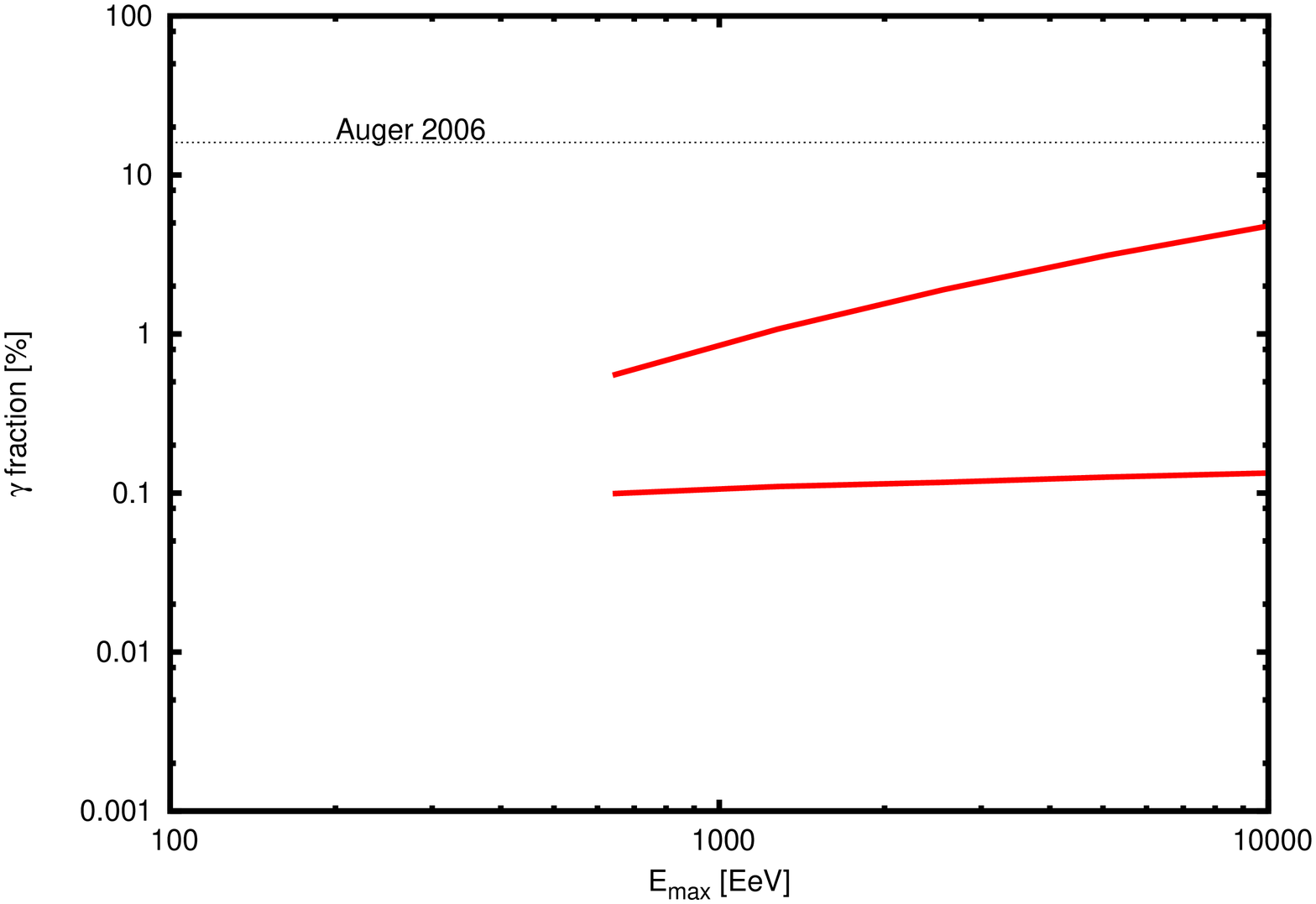}
\includegraphics[width=0.34\textwidth,clip=true,angle=270]{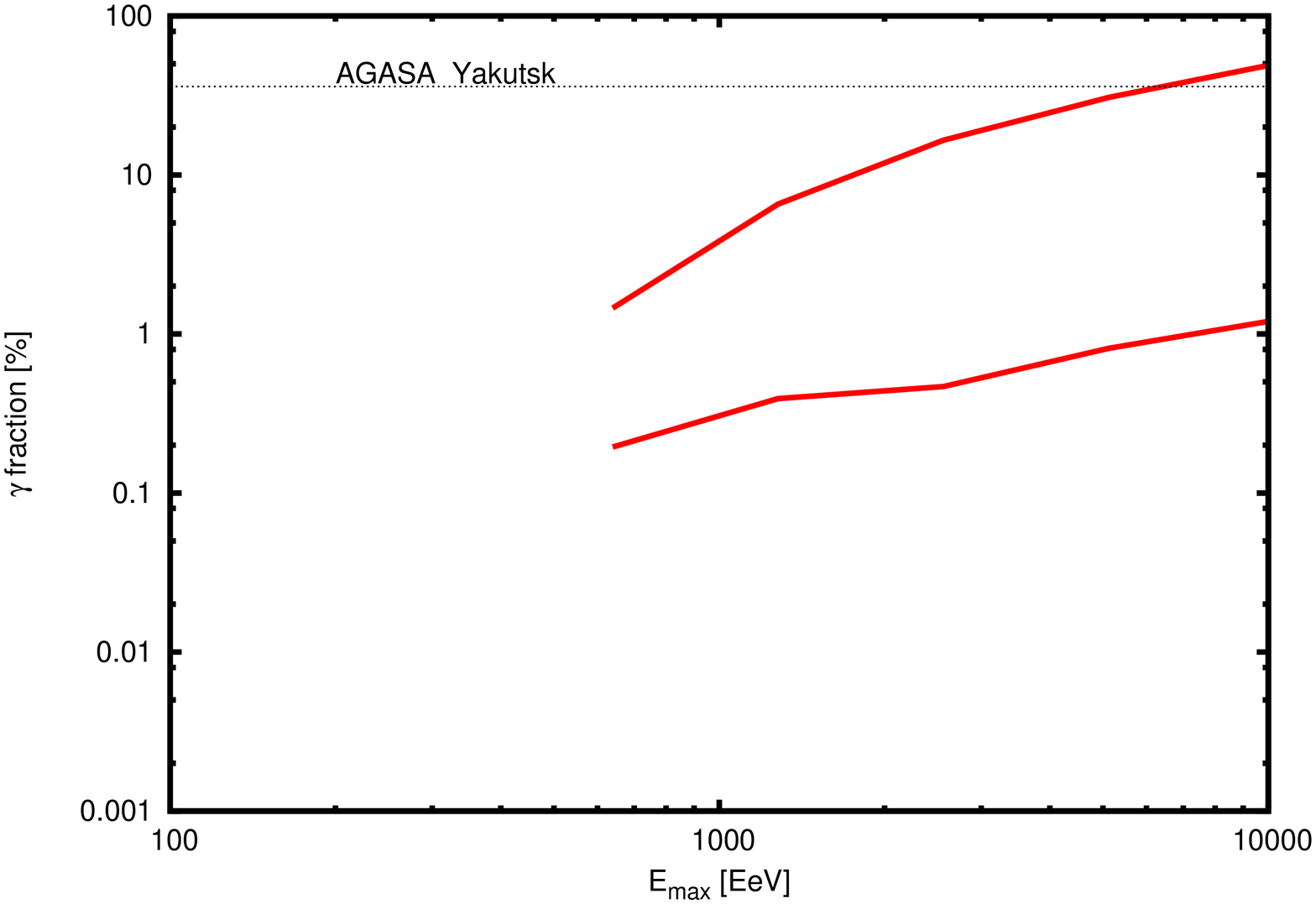}
\includegraphics[width=0.34\textwidth,clip=true,angle=270]{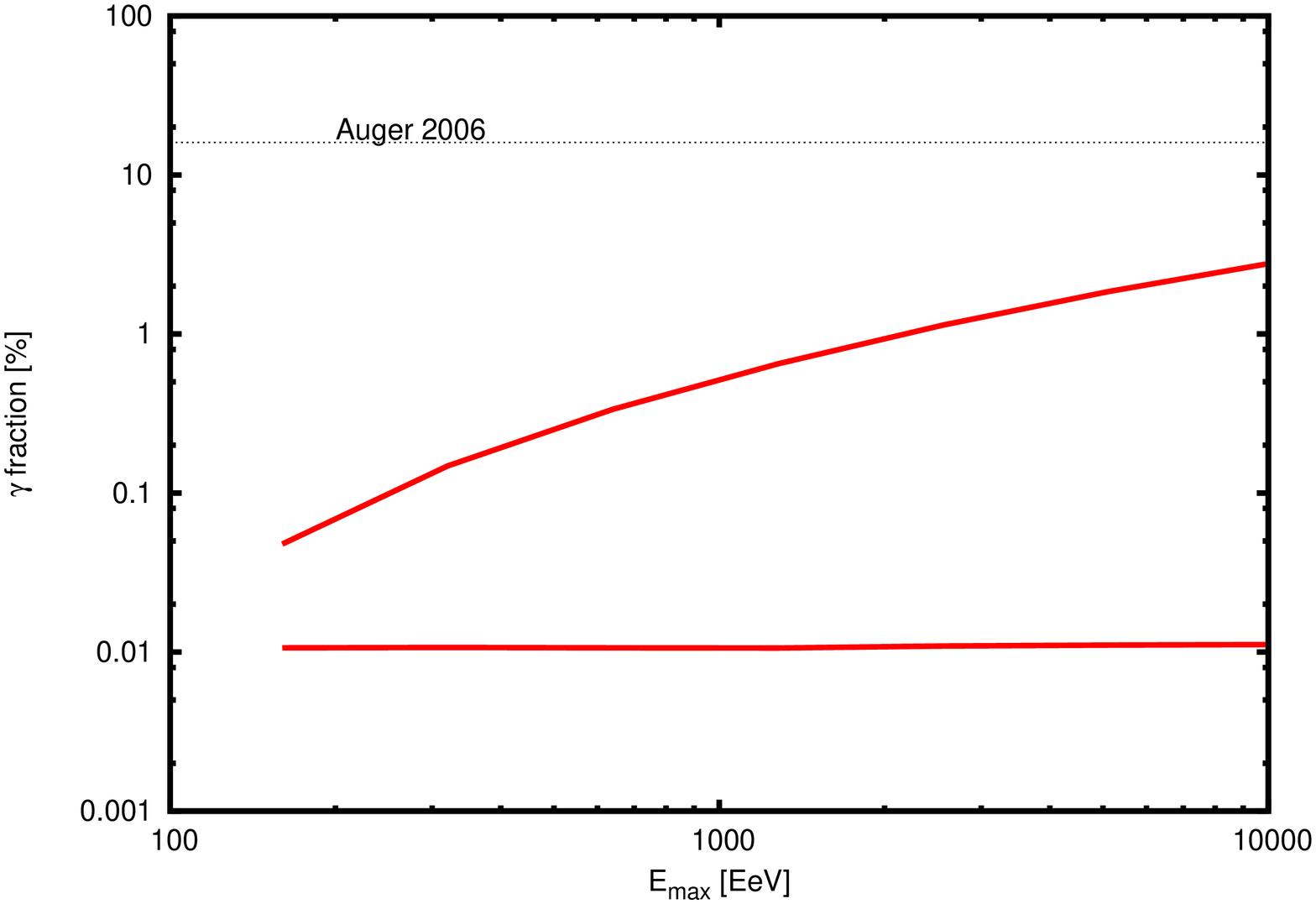}
\includegraphics[width=0.34\textwidth,clip=true,angle=270]{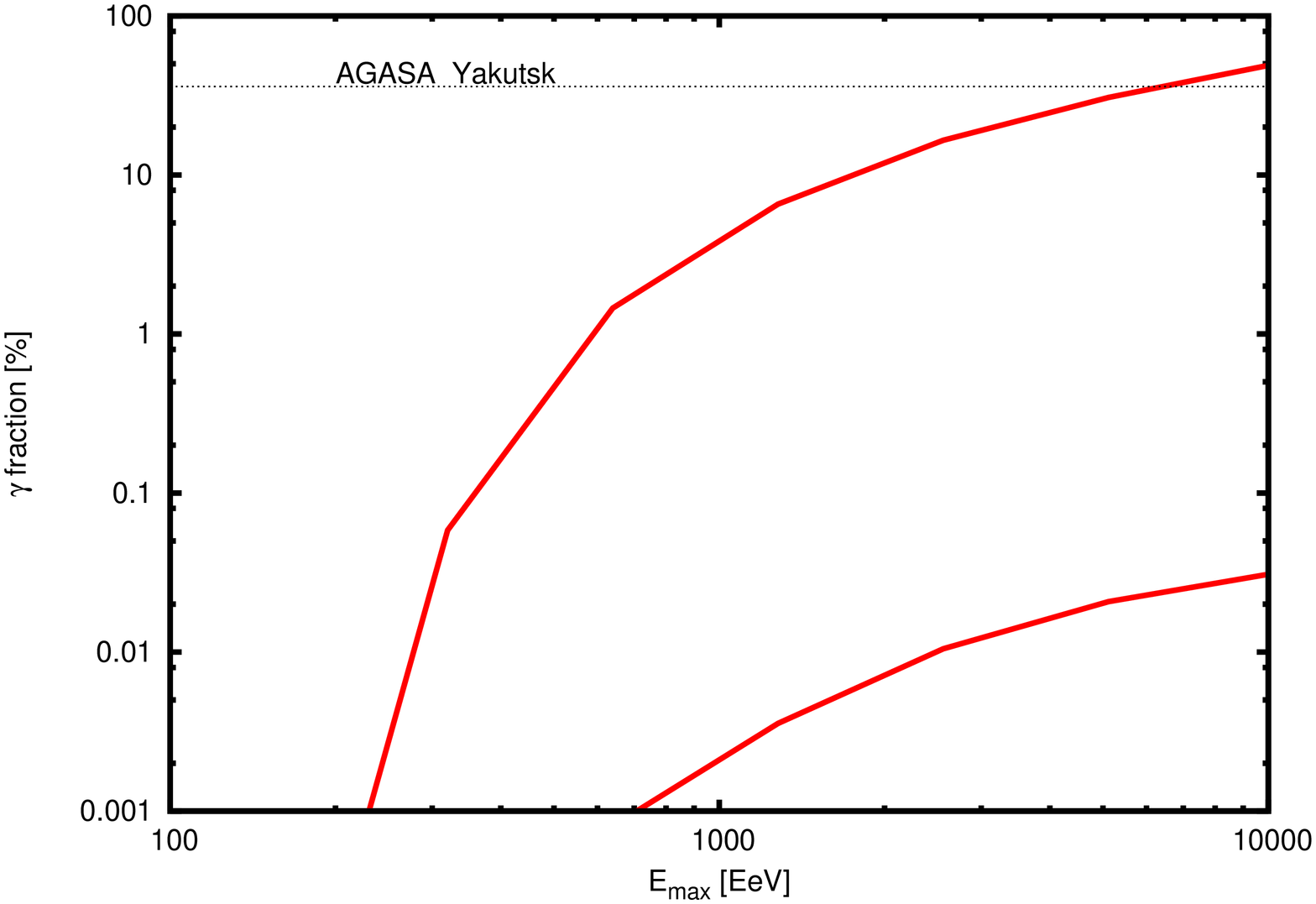}
\caption[...]{Maximum and minimum GZK photon fractions for proton primaries
as  function of $E_{\rm max}$ between $1 \times 10^{20}$~eV and $1.3  \times 10^{21}$~eV, given  in percentage of the integrated fluxes above  $E= 1 \times 10^{19}$~eV (left panels) and $1 \times 10^{20}$eV
(right panels) for AGASA (upper panels) and HiRes (lower panels) respectively, found among the models with $p$-value $>0.05$ in Figs.~\ref{p-values-filter} and \ref{p-values-filter-zmin}. Here we allowed for variable $z_{\rm min} $ between 0 and 0.01.
}
\label{Emax-plots-zmin} 
%Fig. 5  
\end{figure}

So far we have kept the minimum distance to the sources fixed to $z_{\rm min}=0$. Fig.~\ref{p-values-filter-zmin} shows the $p$-values obtained  for $z_{\rm min} = 0.01$ (a minimum distance of 50 Mpc) through the same procedure for the HiRes spectrum. 
 Fig.~\ref{p-values-filter-zmin} shows that the allowed region of models for the HiRes spectrum moves to higher values of $E_{\rm max}$ and $\alpha$ as $z_{\rm min}$ increases. For the AGASA spectrum there are  no allowed solutions within the range of $E_{\rm max}$ and $\alpha$ we consider for $z_{\rm min} = 0.01$ or larger  (the figure would look just black everywhere).

We proceed now to find the maximum and minimum GZK photon fractions.
 First, among all the models with  $z_{\rm min}=$ 0,  0.00125, 0.0025, 0.005, and  0.01 (shown in Fig.~\ref{p-values-filter} and  Fig.~\ref{p-values-filter-zmin} are  only  those with $z_{\rm min}=$ 0 and 0.01 respectively) we chose those with $p$-value$>$ 0.05.  Then, for these models we
compute for a given $E_{\rm max}$ the GZK photon fraction in the  predicted integrated flux above a given energy $E$. Finally,  we choose for each value of $E_{\rm max}$ the values of $\alpha$ and $z_{\rm min}$ for which the GZK photon fraction is either maximum and minimum. 
Figs.~\ref{Emax-plots} and  \ref{Emax-plots-zmin} show the maximum and minimum GZK photon fractions so obtained as a function of $E_{\rm max}$, for 
$z_{\rm min} = 0$, and  variable $z_{\rm min}$ 
 in the range 0 to 0.01, respectively. We give the GZK photon fraction as a percentage
 of the integrated flux above the energy $E$,   for $E= 1 \times 10^{19}$eV (left panels) and $1 \times 10^{20}$eV (right panels) for the AGASA spectrum (upper panels) and HiRes spectrum (lower panels) respectively.  Notice that the 
ranges of GZK fractions do not change much with $z_{\rm min}$ with the exception of the minimum  photon fraction for HiRes above $1 \times 10^{20}$eV, which become much smaller for non zero  $z_{\rm min}$. At  $1 \times 10^{19}$ eV  the photon fractions are always larger than 10$^{-4}$. Figs.~\ref{Emax-plots} and 
Fig.~\ref{Emax-plots-zmin} also shows the experimental upper limits of the photon fraction obtained by Auger in 2006~\cite{Abraham:2006ar} 
 at energies above $1 \times 10^{19}$eV as well as the bound given by  the Yakutsk collaboration combining data from Yakutsk and AGASA, above  $1 \times 10^{20}$~eV~\cite{AgasaYakutskLimit}.

For  comparison, we also include in Fig.~\ref{Emax-plots}  the
 range of photon fractions obtained following the same procedure,  but assuming that either only He or only O or only Fe are emitted by the sources (even though these are not realistic models for the injected composition) and  fitting the UHECR data  with the processed products of photo-disociation of the initial nuclei. The spectrum assumed for the  nuclei is again as  in Eq.~\ref{proton_flux} where $E_{\rm max}$ is now the maximum energy of the  injected nuclei. Here we consider $E_{\rm max}$ between $Z \times 10^{20}$~eV and  $Z\times 1 \times 10^{22}$~eV, where $Z$ is the electric charge of each nucleus. 
 
\begin{figure}[ht]
\includegraphics[height=0.48\textwidth,clip=true,angle=270]{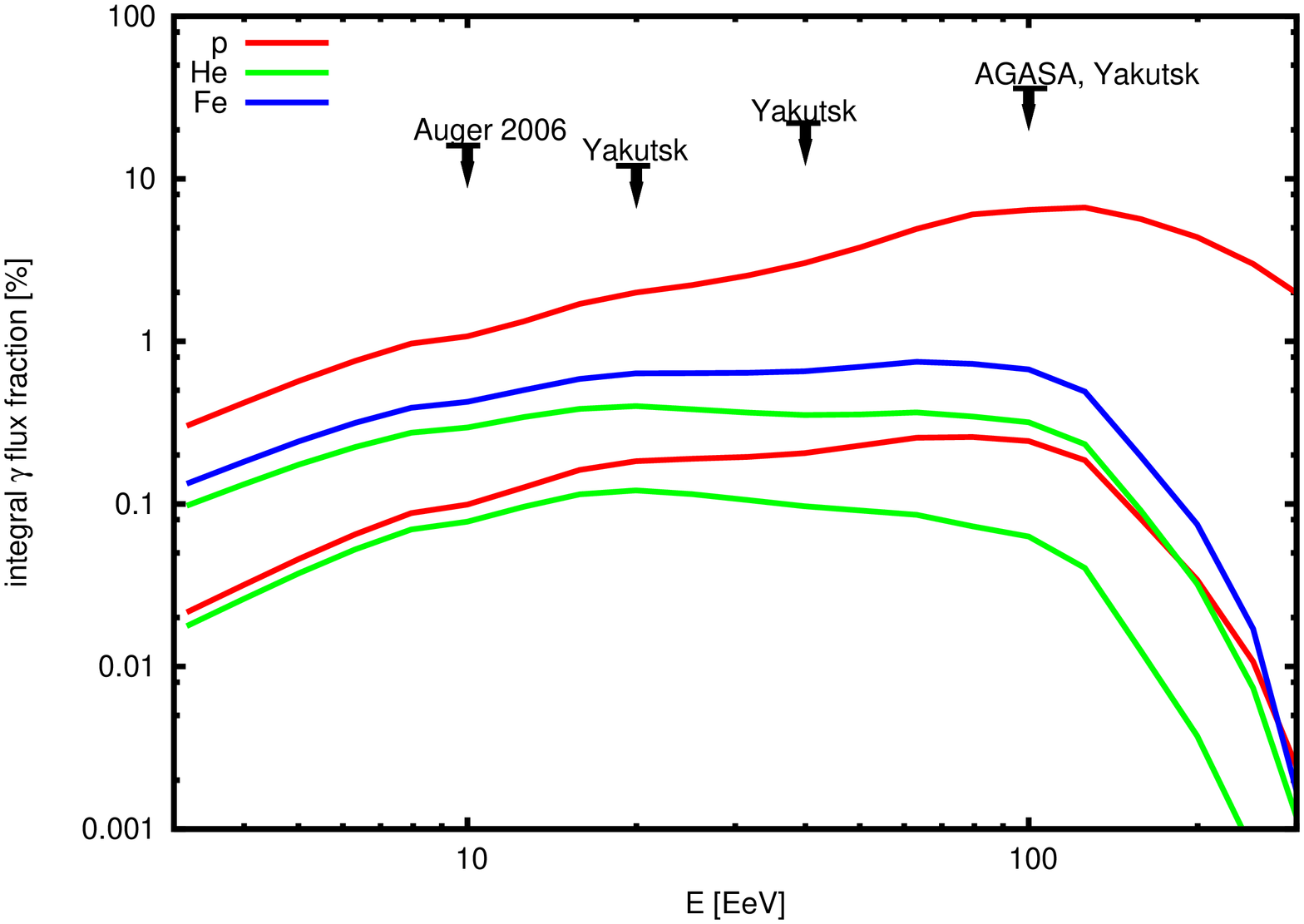}
\includegraphics[height=0.48\textwidth,clip=true,angle=270]{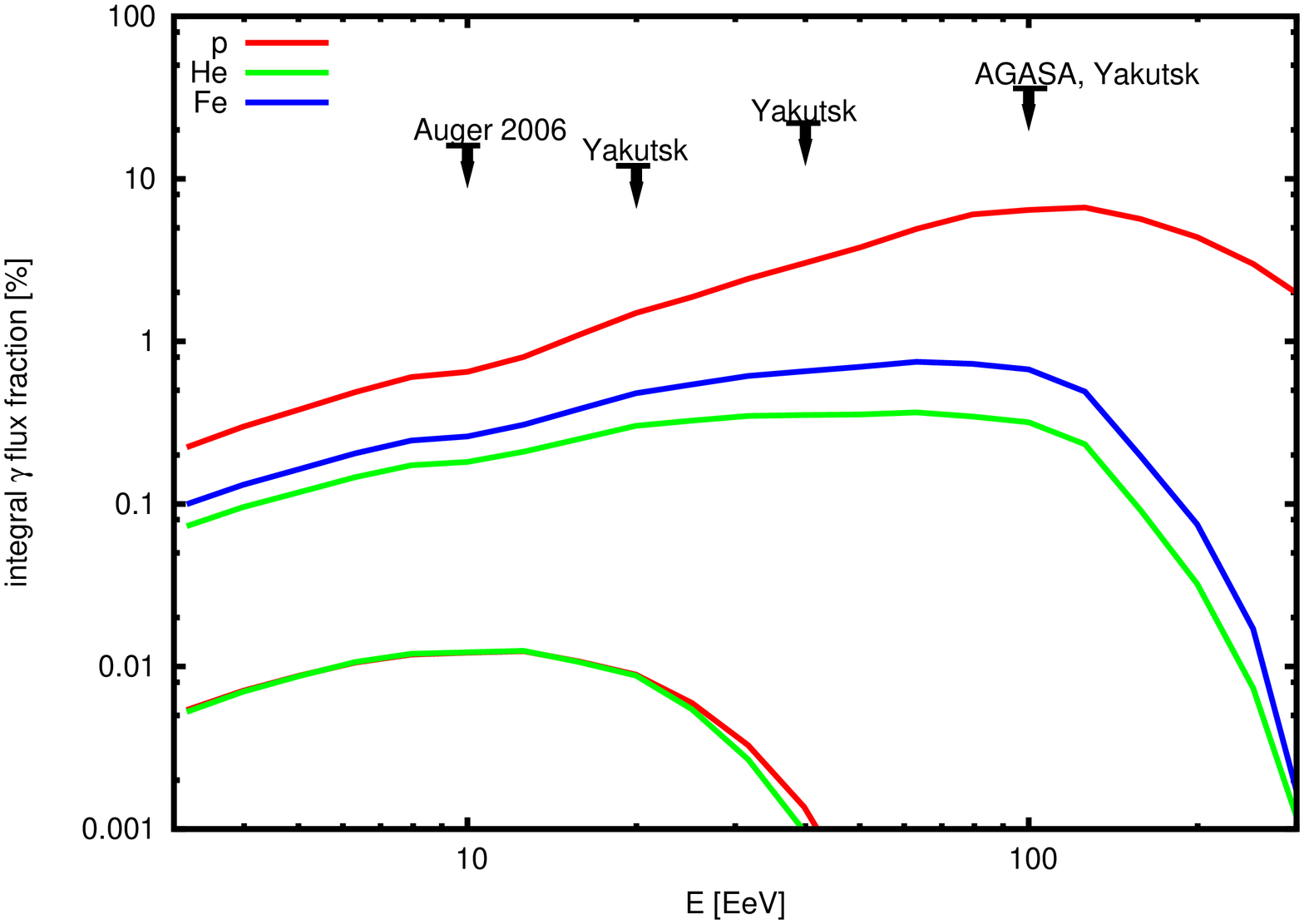}
\caption[...]{Maximum (higher lines) and minimum (lower lines) GZK photon fractions among models with fixed  $z_{\rm min}=0$,  given  as a  percentage of the integrated fluxes above  the energy $E$ are shown as function of $E$ for $E_{\rm max}$ equal to $Z \times1.28 \times 10^{21}$~eV, for AGASA  (left) and HiRes (right). The colors indicate different primaries assumed: red for proton, green for He and blue for Fe. For Fe the minimum fractions are  below the range shown in the figures.}
\label{E-plots} 
%Fig. 6
\end{figure}

\begin{figure}[ht]
\includegraphics[height=0.48\textwidth,clip=true,angle=270]{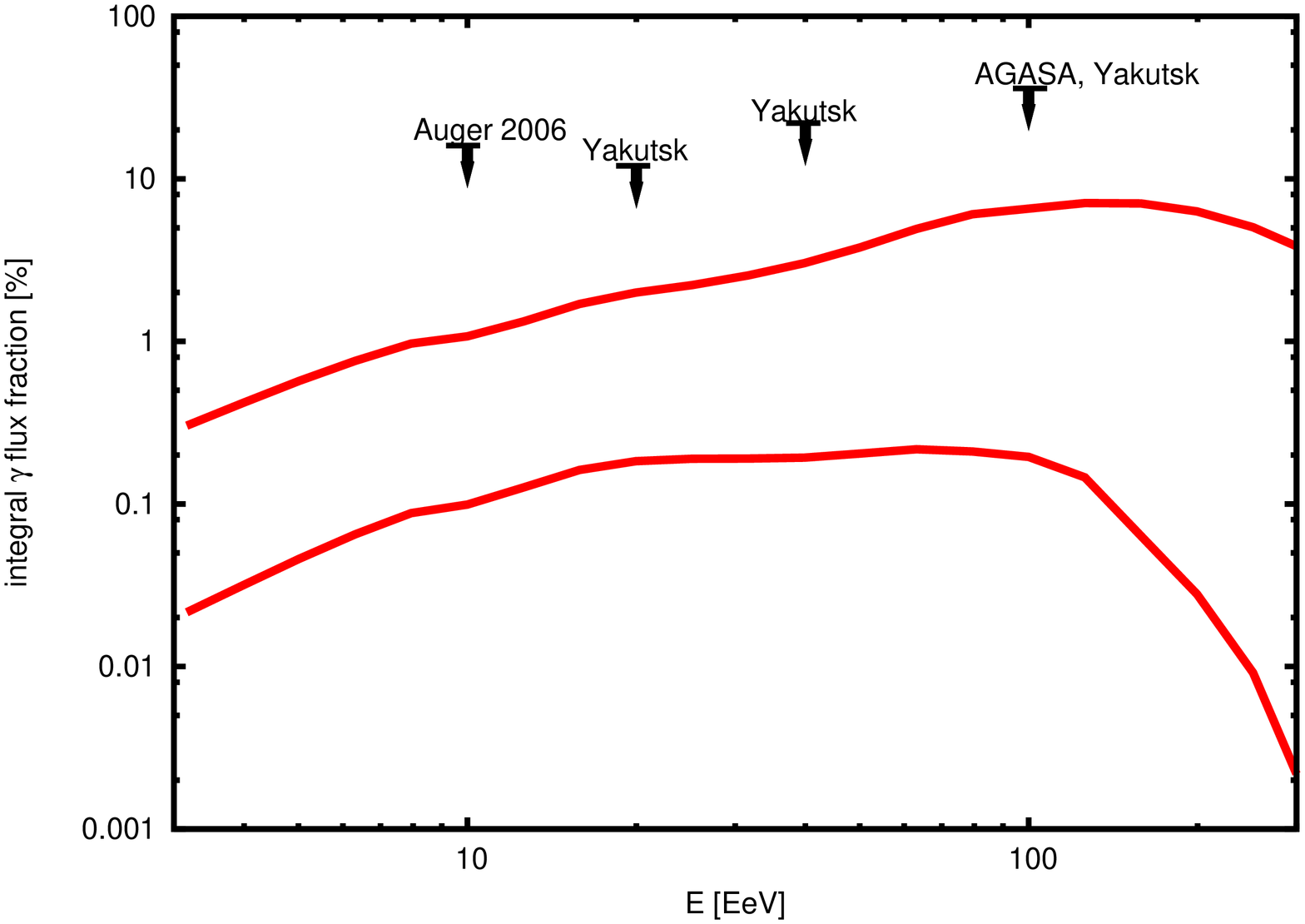}
\includegraphics[height=0.48\textwidth,clip=true,angle=270]{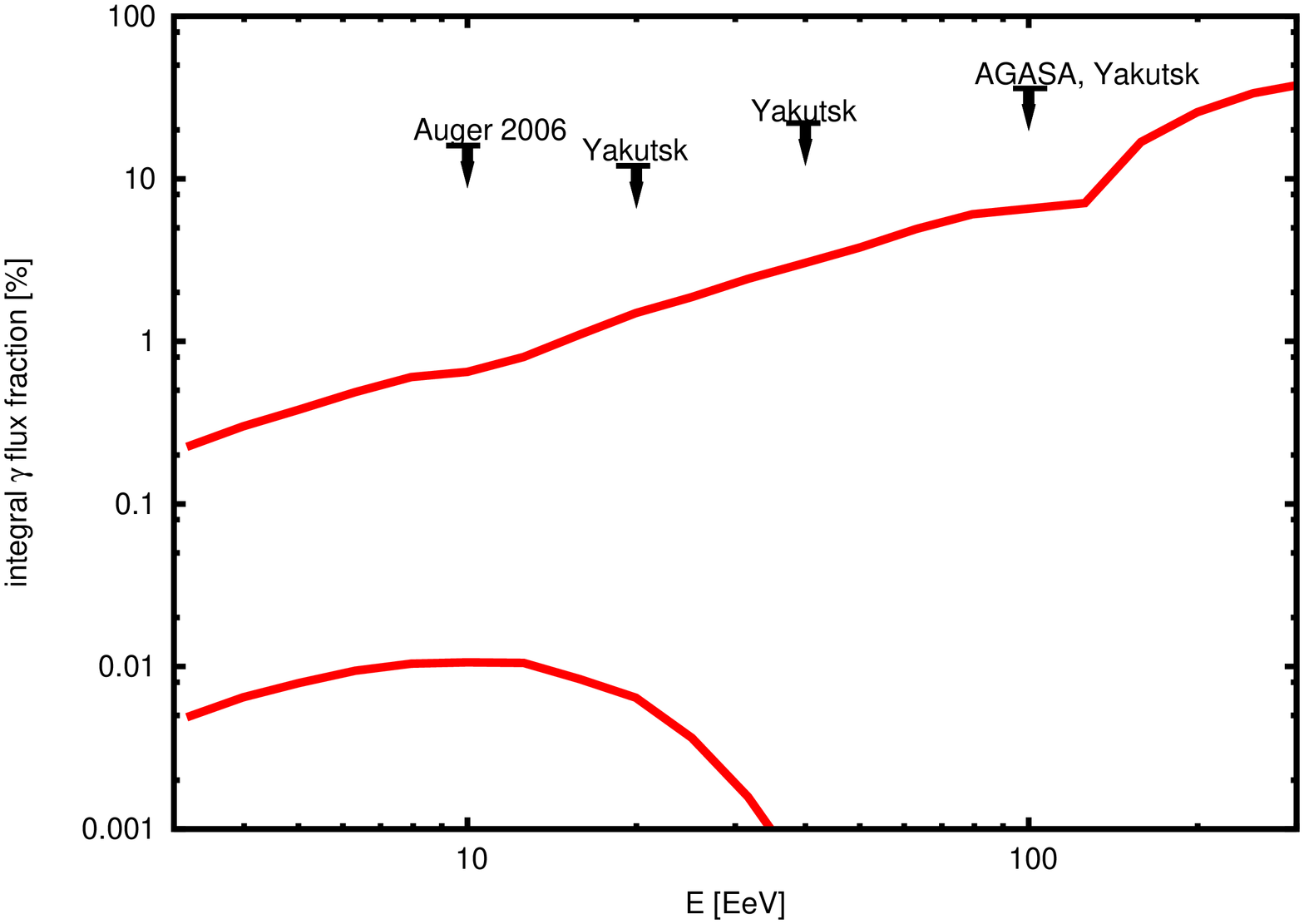}
\caption[...]{Maximum and minimum GZK photon fractions for proton primaries  and  $z_{\rm min}$ varied  between 0 and 0.01. Fractions given  as a percentage of the integrated fluxes above  the energy $E$ are shown as function of $E$  and  $E_{\rm max}$ equal to $1.28 \times 10^{21}$~eV, for AGASA  (left) and HiRes (right).}
\label{E-plots-zmin} 
%Fig. 7 
\end{figure}

In Figs.~\ref{E-plots} and \ref{E-plots-zmin}  the maximum and minimum GZK photon fractions 
in the integrated flux above the energy $E$ are shown as function of $E$ for two values of $E_{\rm max}$,  for fixed $z_{\rm min}=0 $ and for variable $z_{\rm min} $ respectively. The GZK photon fraction is again given  as a percentage of the integrated flux. The ranges of photon fraction do not change much with  $z_{\rm min}$, except for the minimum fraction for HiRes. For  comparison, we also include in Fig.~\ref{E-plots}  the results obtained following the same procedure,  but assuming that either only He or only Fe are emitted by the sources, as explained above.  The photon fractions are given for 
 $E_{\rm max}= Z \times 10^{21}$~eV both for  the AGASA  (left) and  the HiRes (right) spectrum. Upper bounds by Auger~\cite{Abraham:2006ar}, Yakutsk~\cite{Yakutsk} and AGASA-Yakutsk~\cite{AgasaYakutskLimit} are also shown. The maximum  photon fractions  produced by nuclei are not much smaller (within a  factor of 2 to 10 smaller) that those of nucleons, but for Fe the minimum photon fractions can be much smaller (below 10$^{-5}$).

\begin{figure}[ht]
\includegraphics[height=0.48\textwidth,clip=true,angle=270]{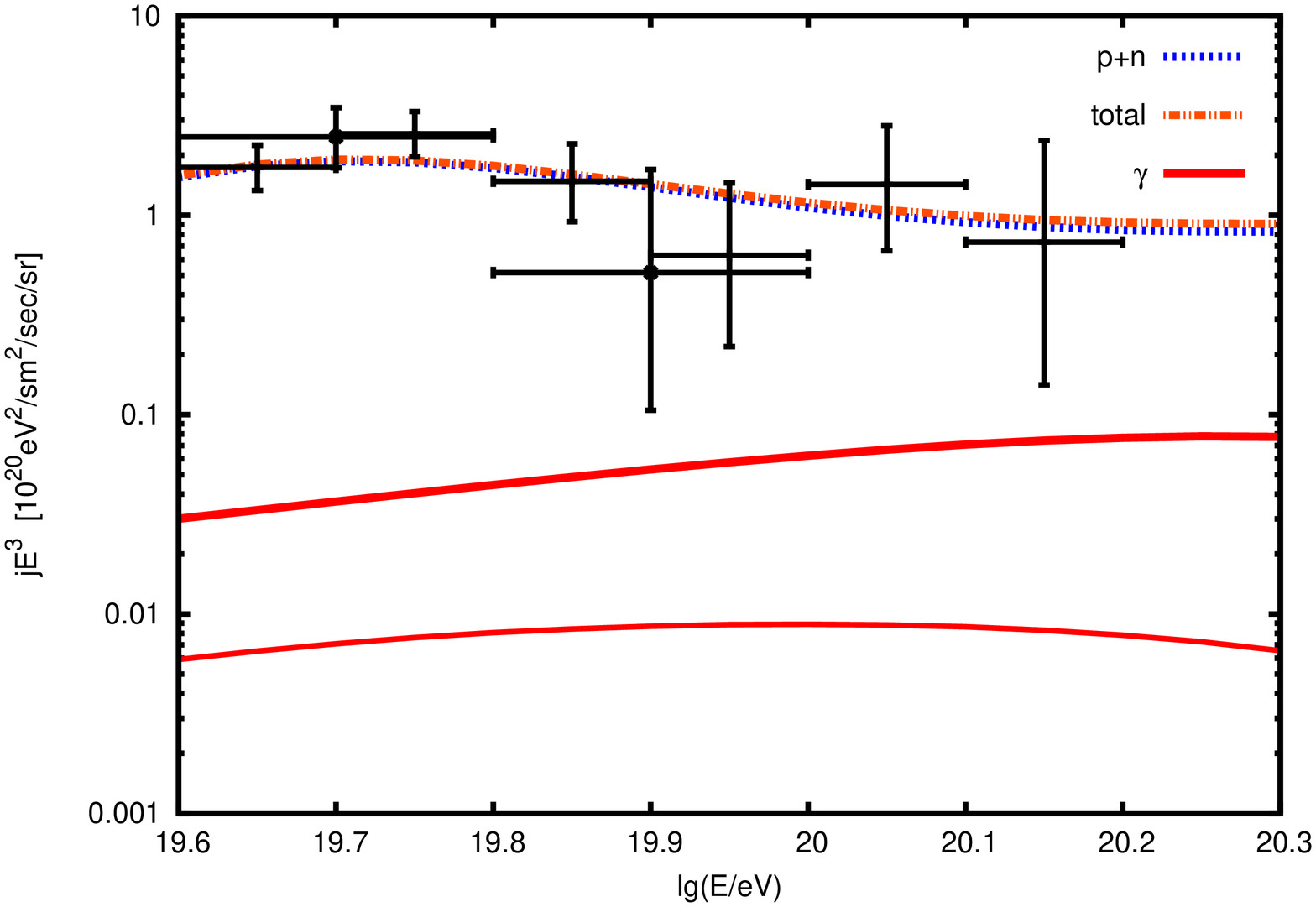}
\includegraphics[height=0.48\textwidth,clip=true,angle=270]{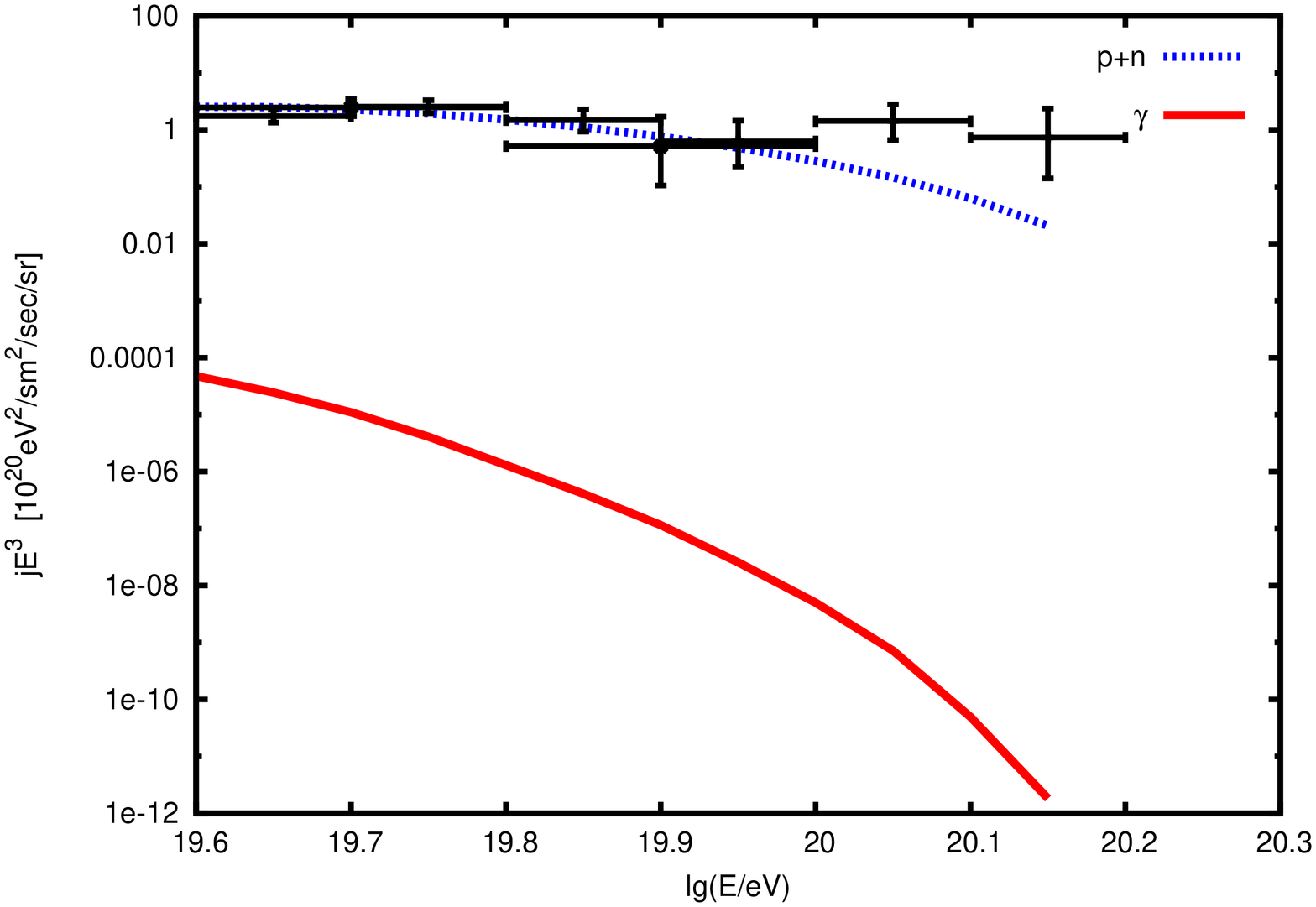}
\caption[...]{Differential  proton, GZK photon  and
total  fluxes which provide good fits  to the  HiRes spectrum for variable $z_{\rm min}$ with either maximal (left panel) or  minimal (right panel) GZK photon content. Left panel: $z_{\rm min}=0.00125$, $m=0$, $E_{\rm max}=1.3 \times 10^{21}$ eV, $\alpha=1$, minimal radio background and $B=10^{-11}$ G for the higher photon line. Lower photon line in the left panel: same source model as higher line, except for maximal intervening radio background and $B=10^{-9}$ G. Right panel: $z_{\rm min}=0.005$, $m=0$, $E_{\rm max}=1.6 \times 10^{20}$ eV, $\alpha=2$, maximal radio background and $B=10^{-9}$ G.}
%Fig 8
\label{HiRes-fits-var-zmin}
\end{figure}

As an example of the fits we obtain to the observed spectra, in  Fig~\ref{HiRes-fits-var-zmin} we show the differential flux of protons, of  GZK photons and the
total differential flux for  two models which provide good fits to   the HiRes UHECR spectrum either with maximal (left panel) or  with minimal (right panel) GZK photon fraction in the integrated flux  above 1 $\times 10^{19}$eV.  In the figures  $z_{\rm min}$ was allowed to vary. 
 The model with maximum GZK photon flux (left panel) has  $z_{\rm min}=0.00125$, $m=0$, $E_{\rm max}=1.3 \times 10^{21}$ eV, $\alpha=1$ and minimal radio background and $B$; the model with minimum photon content  (right panel) has instead $z_{\rm min}=0.005$, $m=0$, $E_{\rm max}=1.6 \times 10^{20}$ eV, $\alpha=2$ and maximal radio background and $B$. In the left panel we show explicitly the effect of the intervening background, given the same source spectrum and distribution. The
lower photon line in the left panel corresponds to the  same source model as the  higher photon line, but   to intervening  radio background and $B$ field that are maximal instead of minimal. We clearly see that the uncertainty in the photon flux due solely to the intervening backgrounds is about one order of magnitude or less in this case. The fit to the HiRes data is the same because photons are in any event subdominant in the flux.


\section{Cosmogenic neutrinos}

The GZK photons and cosmogenic neutrinos are due to the same photo-pion production mechanism:
from the decay of $\pi^0$ we obtain  GZK photons and from the decay of $\pi^{\pm}$ one obtains
neutrinos.  These  ``cosmogenic neutrinos" have been extensively studied, from
1969~\cite{bere} onwards 
(see for example~\cite{reviewGZKneutrinos,Semikoz:2003wv} and
references therein) and constitute one of the main high energy signals
expected in neutrino telescopes, such as ICECUBE~\cite{ICECUBE} 
ANITA~\cite{ANITA} and  SALSA~\cite{SALSA} 
or space based observatories such as EUSO~\cite{EUSO} and
OWL~\cite{OWL}. 

 Thus, GZK photons and cosmogenic neutrinos provide complementary information 
 on the GZK effect. Although they share the same production mechanisn
 GZK photons and cosmogenic neutrinos are affected very differently by intervening backgrounds.
 The flux of GZK photons is affected by the
radio background and EGMF values which do not affect neutrinos. UHE GZK photons only reach us
from less than 100 Mpc away, i.e. a redshift $z< 0.02$. Cosmogenic 
neutrinos do not interact during propagation and thus reach us from 
the whole production volume, which depends on the maximum redshit to the sources
$z_{\rm max}$. Thus the flux of cosmogenic neutrinos arriving to Earth 
depends strongly  on the evolution of the sources, which affect mostly the density of sources far away.
The evolution in comoving volume is here parametrized by the power $m$, so that, excluding the
Hubble expansion, the number density of sources is $\sim (1+z)^{m}$.
\begin{figure}
\includegraphics[width=0.34\textwidth,clip=true,angle=270]{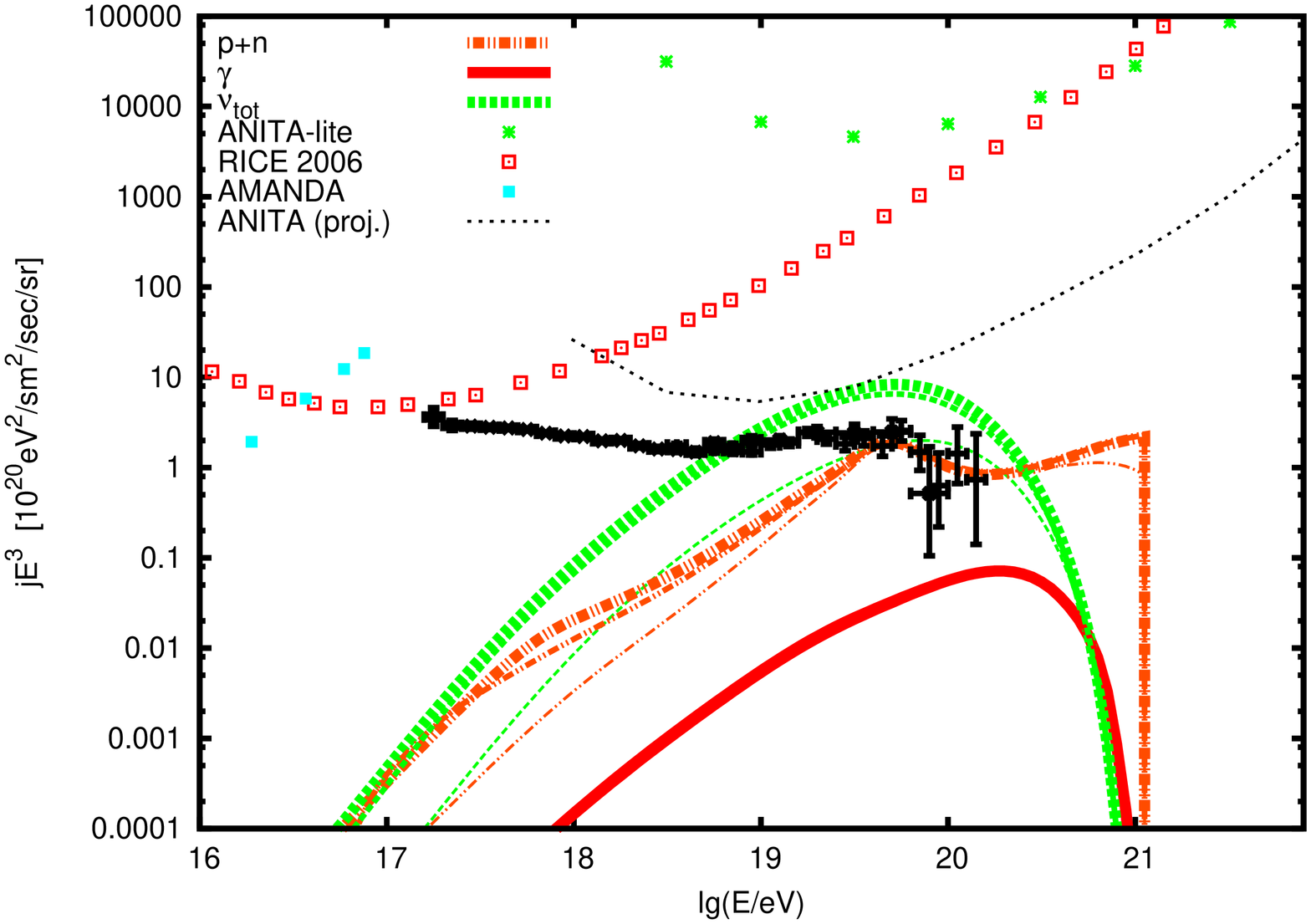}
\includegraphics[width=0.34\textwidth,clip=true,angle=270]{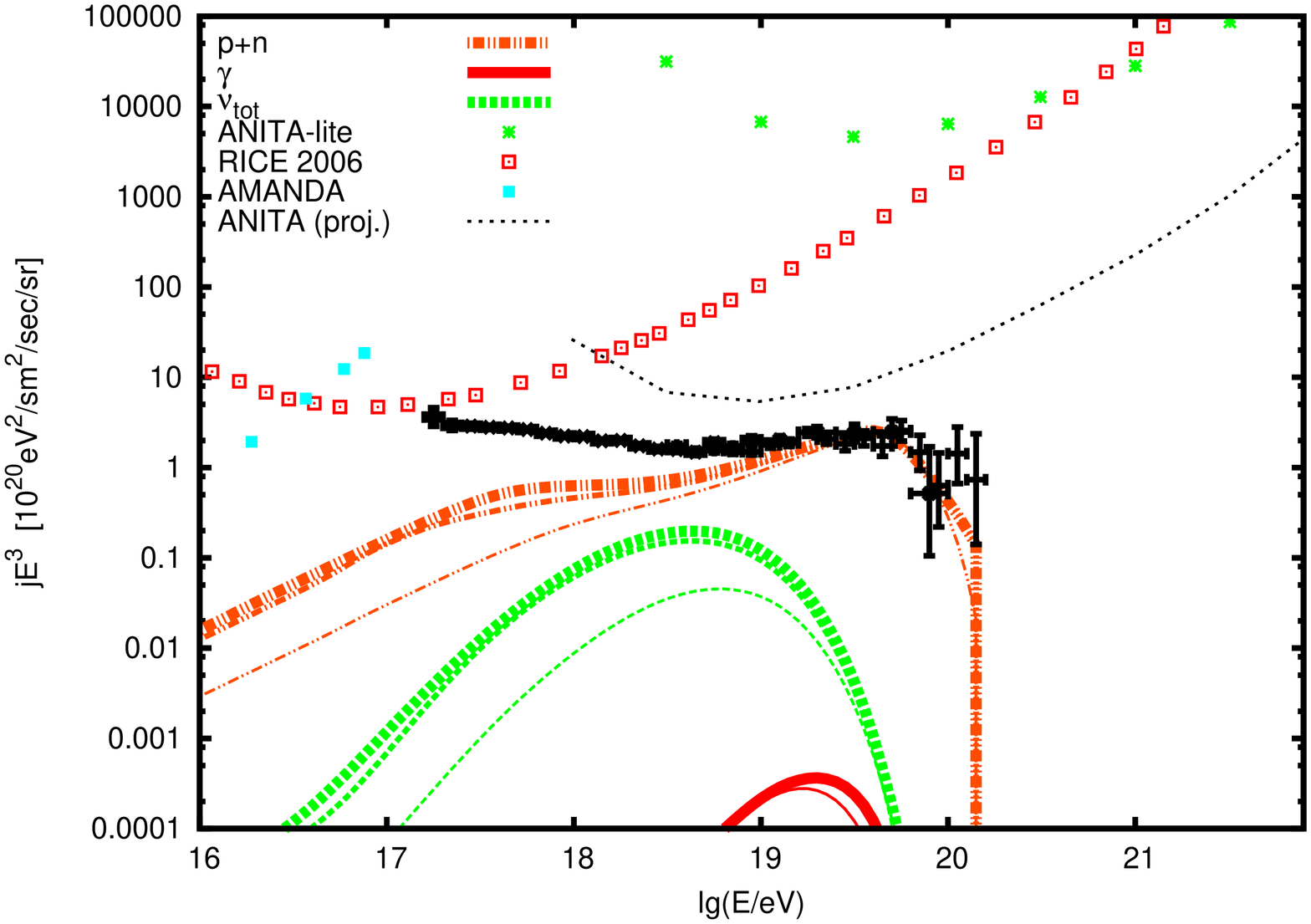}
\caption[...]{Cosmogenic neutrinos for models   which fit the HiRes spectrum with maximum (left panel) and minimum (right panel) GZK photon fractions of  the integrated UHECR fluxes above $E= 1 \times 10^{19}$eV. Same models of Fig.~\ref{HiRes-fits-var-zmin}  but for the three different source evolution models explained in the text.
}
\label{cosmogenic} 
%Fig. 9  
\end{figure}
In Fig.~\ref{cosmogenic} we show six examples of cosmogenic neutrino fluxes for models which fit the HiRes spectrum with either maximum (left panel) or  minimum (right panel) GZK photon fractions of integrated UHECR fluxes above $E= 1 \times 10^{19}$eV, produced by protons emitted at the sources. These are the models of Fig.~\ref{HiRes-fits-var-zmin}  (namely   $E_{\rm max}=1.3 \times 10^{21}$ eV, $\alpha=1$, minimal radio background and $B=10^{-11}$ G  for the left panel and   $E_{\rm max}=1.6 \times 10^{20}$ eV, $\alpha=2$, maximal radio background and $B=10^{-9}$ G for the right panel)   but  assuming three particular source evolution models. The highest neutrino fluxes (thick lines) correspond to the fast star formation rate evolution model of Ref.~\cite{sms05} in which $m=4$ from $z=0$ to $z=1$, then $(1+z)^{m}$ becomes constant (equal to 2$^4=$16) from 
$z=1$ to $z_{\rm max}= 6$ and then goes sharply to zero for $z> 6$. The intermediate neutrino fluxes (thinner dashed lines) correspond to an approximation to the evolution of radio galaxies and AGNs~\cite{dp90}, which is somewhat faster than $m=3$ below $z=1$, peaks at about $z=2$ and decreases rapidly in emissivity at $z>3$. The approximation we used for the figures  has $m=3$ from $z=0$ to $z=1.8$, at which point $(1+z)^{m}$ becomes constant equal to 2.8$^3=$ 22 up to $z_{\rm max}= 3$ where it goes sharply to zero for larger $z$. The lowest neutrinos fluxes in both panels (thinnest dashed lines) correspond to  not  evolving  sources, i.e. $m=0$, and $z_{\rm max}= 3$. The latter is an approximation to an older star population evolution and is taken here as a lower limit to the value of $m$ at low redshits. Negative values of $m$
have been mentioned in the literature only for very massive clusters, which only formed recently. However, accretion shocks in clusters might accelerate heavy nuclei but not protons
 to the energies necessary to account for the ultrahigh energy cosmic rays~\cite{Inoue:2007kn}.
 
 The neutrino flux shown  is the average flux per flavour, that is the total flux of neutrinos and antineutrinos divided by 3. Also shown in the figure are several upper bounds  on cosmogenic neutrinos fluxes by ANITA-Light~\cite{ANITA-Light}, AMANDA~\cite{AMANDA} and  RICE~\cite{RICE} and the  ANITA~\cite{ANITA} projected bound.  

Thick, intermediate and thin lines show the photon (solid red) and baryon p$+$n (dashed double dotted red lines) fluxes for the three source evolutions already mentioned. The baryon flux coincides with the total UHECR predicted, since photon fluxes are always subdominant. 

 Fig.~\ref{cosmogenic} clearly shows that the different source evolution models assumed affect very little (left panel) or not at all (right panel) the GZK photon fluxes but yield different cosmogenic neutrino neutrino fluxes. It is also clear from the figure that the cosmogenic neutrino fluxes are high when GZK photon fluxes are high and vice versa. We have not attempted here to maximize or minimize the cosmogenic neutrino fluxes, just to provide  some examples of the expected range of fluxes given a particular GZK photon flux. Moreover, since here we fit the UHECR spectrum only above $4 \times 10^{19}$~eV, we predict accurately the neutrino spectrum at energies above $4 \times 10^{18}$. At lower energies higher neutrino fluxes could be predicted if there is an extragalactic component to the UHECR at energies below $4 \times 10^{19}$~eV. 

\section{Conclusions}

 The Pierre Auger Observatory has already set bounds on the photon fraction above 10$^{19}$~eV~\cite{Abraham:2006ar} and better bounds are expected soon.  We expect to have in the near future the high statistic data that may allow to study a subdominant component of
UHECR consisting of photons.  
Auger will hopefully see photons or place a limit 
on the photon fraction  at the level of  a few \% or below~\cite{Risse:2007sd}. If complemented by an extended northern array,
a sensitivity level of below 0.1\% could be reached within a few years of full 
operation~\cite{Risse:2007sd}.
In this paper we  address the physical implications of such detection
or limits.  

Here we calculate the flux of ``GZK-photons", namely the flux of 
photons produced by  extragalactic nucleons through the resonant
 photoproduction of pions, the  GZK effect.  This flux  depends on the
UHECR spectrum on Earth, of the spectrum of nucleons emitted at the sources, which we characterize by its slope  and maximum energy, on the distribution of
sources and on  the intervening cosmological backgrounds, in particular the magnetic field and radio backgrounds.  
We compute the possible range of
the GZK photon fraction of the total UHECR flux 
 for the AGASA   and the HiRes spectra.
We fit the UHECR data  above  4 $\times 10^{19}$eV either
 minimizing or maximizing the number of GZK protons produced. We find
(see Figs.~\ref{E-plots} and \ref{E-plots-zmin}) that  assuming exclusively
nucleons are emitted at the sources the GZK photon fraction of the total 
integrated  UHECR flux  could reach a few \% above 10$^{19}$~eV  and  10\%
 above 10$^{20}$~eV, or be between one (for AGASA) and several (for HiRes) orders of magnitude smaller,  under the level that could be detected at Auger South. The maximum photon fractions do not depend much from the minimum distance to the sources. 
 
 We find (as in  Ref.~\cite{Gelmini:2007sf}) that the  photon fraction in cosmic rays at
energies above $10^{19}$ eV could be as low as $10^{-4}$ (Ref.~\cite{Sigl:2007ea} finds comparable small fractions). Photon fluxes so small could only be detected in future experiments like Auger North~\cite{Auger_North} plus South~\cite{Risse:2007sd}, EUSO~\cite{EUSO}  and OWL~\cite{OWL}.

Just for comparison in  Fig.~\ref{E-plots}  we also show  the range of GZK-photon 
fractions expected if purely He or purely Fe nuclei were emitted at the sources, 
 and in both cases the maximum GZK photon fractions expected are smaller by a factor between 2 and 10. For nuclei produced at the sources the maximum photon fraction is a factor of 2 to 3 times smaller above 10$^{19}$~eV but the minimum could be much smaller than for nucleons.

The detection of UHECR photons would open a new window for ultra-high energy astronomy and   help  
establish the UHECR sources. The detection of a larger photon flux than expected for GZK photons given the particular UHECR spectrum,
would imply the emission of photons at the
source or new physics, such as Top-Down models. If photons are not seeing,
 Auger will  place  interesting bounds
on GZK photon production models.

 Finally we also briefly  comment on cosmogenic neutrino fluxes, which provide complementary information on the GZK effect. Although they share the same production mechanisn,
 GZK photons are affected by the
radio background and EGMF  which do not affect neutrinos and, contrary to photons, cosmogenic 
neutrinos depend strongly on the evolution of sources. In any event, as shown in  Fig.~\ref{cosmogenic},
cosmogenic neutrino fluxes are high when GZK photon fluxes are high and vice versa.

\verb''\ack
%\vspace{0.3cm}
%{\bf Acknowledgments}

The work of G.G and O.K.
 was supported in part by NASA grants NAG5-13399 and ATP03-0000-0057. 
G.G was also supported in part by the US DOE grant DE-FG03-91ER40662
Task C. The numerical part of this work was performed at the computer cluster of the   INR RAS Theory Division and  the ``Neutrino" cluster of  the UCLA Physics and Astronomy Department.


\vspace{0.3cm}
\begin{thebibliography}{99}

\bibitem{gzk}
K.~Greisen,
%``End To The Cosmic Ray Spectrum?,''
Phys.\ Rev.\ Lett.\  {\bf 16}, 748 (1966).
%%CITATION = PRLTA,16,748;%%
G.~T.~Zatsepin and V.~A.~Kuzmin,
%``Upper Limit Of The Spectrum Of Cosmic Rays,''
JETP Lett.\  {\bf 4}, 78 (1966)
[Pisma Zh.\ Eksp.\ Teor.\ Fiz.\  {\bf 4}, 114 (1966)].
%CITATION = JTPLA,4,78;%%
%ref 1


\bibitem{agasa}
M.~Takeda {\it et al.},
%``Extension of the cosmic-ray energy spectrum beyond the predicted 
% Greisen-Zatsepin-Kuzmin cutoff,''
Phys.\ Rev.\ Lett.\  {\bf 81}, 1163 (1998);
%CITATION = ASTRO-PH 9807193;%%
see N.~Hayashida {\it et al.},
%``Updated AGASA event list above 4*10**19-eV,''
astro-ph/0008102,
%CITATION = ASTRO-PH 0008102;%%
for an update;  see also
{\sf http~://www-akeno.icrr.u-tokyo.ac.jp/AGASA/}.
%ref 2

%\cite{agasa_spec}
\bibitem{agasa_spec}
  M.~Takeda {\it et al.},
  %``Energy determination in the Akeno Giant Air Shower Array experiment,''
  Astropart.\ Phys.\  {\bf 19}, 447 (2003)
  [astro-ph/0209422]. 
  %%CITATION = ASTRO-PH 0209422;%
  %ref 3 
 

\bibitem{hires}
R.~U.~Abbasi {\it et al.}  [High Resolution Fly's Eye Collaboration],
%``Measurement of the flux of ultrahigh energy cosmic rays from monocular
%observations by the High Resolution Fly's Eye experiment,''
Phys.\ Rev.\ Lett.\  {\bf 92}, 151101 (2004);
%%CITATION = ASTRO-PH 0208243;%%
see also {\sf http~://hires.physics.utah.edu/}.
%ref 4

%\cite{hires_mono_spec}
\bibitem{hires_mono_spec}
R.~Abbasi {\it et al.}  [HiRes Collaboration],
  %``Observation of the GZK cutoff by the HiRes experiment,''
  astro-ph/0703099.
%ref 5

\bibitem{50Mpc} F.W. Stecker, Phys. Lett. {\bf 21}, 1016 (1968); S.~Yoshida and M.~Teshima, 
Prog. Theor. Phys. {\bf 89}, 833 (1993); F.~A.~Aharonian and J.~W.~Cronin, {Phys. Rev.}
{\bf D50}, 1892 (1994); J.~W.~Elbert and P.~Sommers,
{Astrophys. J.} {\bf 441}, 151 (1995);
%ref 6

\bibitem{40Mpc} F.~Halzen, R.~A.~Vazquez, T.~Stanev, and V.~P.~Vankov,
  Astropart. Phys., {\bf 3}, 151 (1995).
  %ref 7

\bibitem{dolag2004}
K.~Dolag, D.~Grasso, V.~Springel and I.~Tkachev,
%``Mapping deflections of Ultra-High Energy Cosmic Rays in Constrained
%Simulations of Extragalactic Magnetic Fields,''
JETP Lett.\  {\bf 79}, 583 (2004)
[Pisma Zh.\ Eksp.\ Teor.\ Fiz.\  {\bf 79}, 719 (2004)]; and
%%CITATION = ASTRO-PH 0310902;%%
%K.~Dolag, D.~Grasso, V.~Springel and I.~Tkachev,
%``Constrained simulations of the magnetic field in the local universe and the
%propagation of UHECRs,''
JCAP {\bf 0501}, 009 (2005).
%%CITATION = ASTRO-PH 0410419;%%
%ref 8

\bibitem{Sigl:2004yk}
G.~Sigl, F.~Miniati and T.~A.~Ensslin,
%``Ultra-high energy cosmic rays in a structured and magnetized universe,''
Phys.\ Rev.\ D {\bf 68}, 043002 (2003);
%%CITATION = ASTRO-PH 0302388;%%
%
%``Signatures of magnetized large scale structure in ultra-high energy  cosmic
%rays,''
astro-ph/0309695;
%%CITATION = ASTRO-PH 0309695;%%
%
%``Ultra-high energy cosmic ray probes of large scale structure and magnetic
%fields,''
Phys.\ Rev.\ D {\bf 70}, 043007 (2004);
%%CITATION = ASTRO-PH 0401084;%%
%``Cosmic magnetic fields and their influence on ultra-high energy cosmic ray
%propagation,''
astro-ph/0409098.
%%CITATION = ASTRO-PH 0409098;%%
%ref 9

\bibitem{berezinsky2002}
  V.~Berezinsky, A.~Z.~Gazizov and S.~I.~Grigorieva,
  %``On astrophysical solution to ultra high energy cosmic rays,''
  hep-ph/0204357.
  %%CITATION = HEP-PH 0204357;%%
  %ref 10

\bibitem{galactic_magn_field}
 T. T.~Stanev,
  %``Ultra high energy cosmic rays and the large scale structure of the
  %galactic magnetic field,''
  Astrophys.\ J.\  {\bf 479}, 290 (1997);
 % [astro-ph/9607086];
  %%CITATION = ASTRO-PH 9607086;%%
  G.~A.~Medina-Tanco, E.~M.~de Gouveia Dal Pino and J.~E.~Horvath,
  %``Deflection of ultra high energy cosmic rays by the galactic magnetic
  %field: From the sources to the detector,''
  astro-ph/9707041;
  %%CITATION = ASTRO-PH 9707041;%%
  M.~Prouza and R.~Smida,
  %``The Galactic magnetic field and propagation of ultra-high energy cosmic
  %rays,''
  astro-ph/0307165.
  %%CITATION = ASTRO-PH 0307165;%%
  %ref 11

\bibitem{Auger} Pierre Auger Observatory, http://www.auger.org.
%ref 12


\bibitem{bere} V. S. Berezinsky and G. T. Zatsepin, Phys. 
Lett {\bf 28B}, 423 (1969).
%ref 13

\bibitem{reviewGZKneutrinos}
O.~E.~Kalashev, V.~A.~Kuzmin, D.~V.~Semikoz and G.~Sigl,
%``Ultra-high energy neutrino fluxes and their constraints,''
Phys.\ Rev.\ D {\bf 66}, 063004 (2002).
%[arXiv:hep-ph/0205050].
%ref 14


\bibitem{Semikoz:2003wv}
D.~V.~Semikoz and G.~Sigl,
%``Ultra-high energy neutrino fluxes: New constraints and implications,''
JCAP {\bf 0404}, 003 (2004).
%[hep-ph/0309328].
%ref 15

\bibitem{ICECUBE} ICECUBE collaboration, http://icecube.wis.edu/.
%ref 16


\bibitem{ANITA} ANtarctic Impulse Transient Array (ANITA),
  http://www.ps.uci.edu/\~{}anita/.
  %ref 17
  

\bibitem{SALSA} Saltdome Shower Array,
P.~Gorham {\em et al.},  Nucl.\ Instrum.\ Meth.\ A {\bf 490}, 476 (2002).
%ref 18


\bibitem{EUSO} Extreme Universe Space Observatory,
{\it http://www.euso-mission.org/}. ``JEM/EUSO Project"
%: To Study Extreme Universe by Wide-angle Telescope''
Talk by Inoue, I.; Ebisuzaki, E. on 6th COSPAR Scientific Assembly. 
Held 16 - 23 July 2006, in Beijing, China., p.2902. 
%ref 19

\bibitem{OWL} Orbiting Wide-angle Light-collectors,
{\it http://owl.gsfc.nasa.gov/}.  F.~W.~Stecker {\em et al.}
  %``Observing the ultrahigh energy universe with OWL eyes,''
  Nucl.\ Phys.\ Proc.\ Suppl.\  {\bf 136C}, 433 (2004).
 % [arXiv:astro-ph/0408162].
  %%CITATION = NUPHZ,136C,433;%%
  %ref 20


\bibitem{wdowczyk}
J. Wdowczyk , W. Tkaczyk, C. Adcock and A. W. Wolfendale, 
J. Phys. A: Gen. Phys {\bf 4}
L37-9 (1971); J. Wdowczyk , W. Tkaczyk and A. W. Wolfendale,
 J. Phys. A: Gen. Phys {\bf 5}
1419 (1972); J. Wdowczyk  and A. W. Wolfendale, Astrophys. 
Jour. {\bf 349}, 35 (1990)
%ref 21

\bibitem{Aharonian1990}
F.A.~Aharonian, V.V.~Vardanian and B.L.~Kanevsky,
Astrophysics and Space Science {\bf 167}, 111 (1990);
%Corresponding proton GZK cutoff calculation was done in
{\it ibid} {\bf 93}, 111 (1990).
%ref 22



\bibitem{SiglOlinto95}
  S.~j.~Lee, A.~Olinto and G.~Sigl,
  %``Extragalactic magnetic field and the highest energy cosmic rays,''
  Astrophys.\ J.\  {\bf 455}, L21 (1995).
 % [astro-ph/9508088].
  %%CITATION = ASTRO-PH 9508088;%%
  %ref 23

\bibitem{astro_photons}
O.~E.~Kalashev, V.~A.~Kuzmin, D.~V.~Semikoz and I.~I.~Tkachev,
%``Photons as ultra high energy cosmic rays?,''
astro-ph/0107130.
%%CITATION = ASTRO-PH 0107130;%%
%ref 24

\bibitem{Tinyakov:2001nr}
  P.~G.~Tinyakov and I.~I.~Tkachev,
  %``BL Lacertae are sources of the observed ultra-high energy cosmic rays,''
  JETP Lett.\  {\bf 74}, 445 (2001)
  [Pisma Zh.\ Eksp.\ Teor.\ Fiz.\  {\bf 74}, 499 (2001)].
 % [astro-ph/0102476].
  %%CITATION = ASTRO-PH 0102476;%%
  %ref 25
  
\bibitem{gzk_photon} 
G.~Gelmini, O.~Kalashev and D.~V.~Semikoz,
% ``GZK  photons as ultra high energy cosmic rays,'' 
 astro-ph/0506128.
  %%CITATION = ASTRO-PH 0506128;%%
%ref26

%\cite{Gelmini:2007sf}
\bibitem{Gelmini:2007sf}
  G.~Gelmini, O.~Kalashev and D.~V.~Semikoz,
  %``GZK photons in the minimal ultra high energy cosmic rays model,''
  astro-ph/0702464.
  %%CITATION = ASTRO-PH/0702464;%%
  %ref 27

\bibitem{bertoux}
X. Bertou, P. Billoir, and S. Dagoret-Campagne,
%``LPM effect and pair production in the geomagnetic field: a 
% signature of ultra-high energy photons in the Pierre Auger Observatory,"
Astropart.\ Phys.\  {\bf 14}, 121 (2000).
%ref 28

%\cite{Risse:2007sd}
\bibitem{Risse:2007sd}
  M.~Risse and P.~Homola,
  %``Search for ultra-high energy photons using air showers,''
  Mod.\ Phys.\ Lett.\  A {\bf 22}, 749 (2007).
 % [arXiv:astro-ph/0702632].
 %ref 29

%\cite{Abraham:2006ar}
\bibitem{Abraham:2006ar}
   J.~Abraham {\it et al.}  [Pierre Auger Collaboration],
  %``An upper limit to the photon fraction in cosmic rays above 10**19-eV  from
  %the Pierre Auger observatory,''
  Astropart.\ Phys.\  {\bf 27}, 155 (2007).
%  [arXiv:astro-ph/0606619].
%ref 30



\bibitem{Auger_North}
Auger North web-site can be found here:
{\it http://www.augernorth.org/}.
%ref 31
 
 %\cite{Sigl:2007ea}
\bibitem{Sigl:2007ea}
  G.~Sigl,
  %``Non-Universal Spectra of Ultra-High Energy Cosmic Ray Primaries and
  %Secondaries in a Structured Universe,''
  Phys.\ Rev.\  D {\bf 75}, 103001 (2007).
%  [arXiv:astro-ph/0703403]. 
%ref 32

\bibitem{kks1999}
O.E.~Kalashev, V.A.~Kuzmin and D.V.~Semikoz,
%``Top-down models and extremely high energy cosmic rays,''
astro-ph/9911035.
%%CITATION = ASTRO-PH 9911035;%% 
%O.~E.~Kalashev, V.~A.~Kuzmin and D.~V.~Semikoz,
  %``Ultra high energy cosmic rays propagation in the galaxy and  anisotropy,''
  Mod.\ Phys.\ Lett.\ A {\bf 16}, 2505 (2001);
%[astro-ph/0006349].
%%CITATION = ASTRO-PH 0006349;%%
O.E.~Kalashev Ph.D. Thesis, INR RAS, 2003.
%ref 33

\bibitem{reviews1} P.~Bhattacharjee, G.~Sigl, Phys. Rept. {\bf 327},
109 (2000).
%ref 34

\bibitem{propagLeeSigl}
  S.~Lee,
  %``On the propagation of extragalactic high-energy cosmic and gamma-rays,''
  Phys.\ Rev.\ D {\bf 58}, 043004 (1998);
  %[astro-ph/9604098].
  %%CITATION = ASTRO-PH 9604098;%%
    G.~Sigl, S.~Lee and P.~Coppi,
  %``The universal gamma-ray flux, grand unified theories, and extragalactic
  %magnetic field,''
  astro-ph/9604093.
  %%CITATION = ASTRO-PH 9604093;%%
  %ref 35

%\cite{Mucke:1999yb}
\bibitem{Mucke:1999yb}
  A.~Mucke, R.~Engel, J.~P.~Rachen, R.~J.~Protheroe and T.~Stanev,
 %  ``Monte Carlo simulations of photohadronic processes in astrophysics,''
  Comput.\ Phys.\ Commun.\  {\bf 124}, 290 (2000).
 % [astro-ph/9903478].
  %%CITATION = ASTRO-PH 9903478;%%
  %ref 36

\bibitem{puget} J.L. Puget, F. W. Stecker and J. Bredekamp
Astrophys.  J. {\bf 205}, 638 (1976).
%ref 37


%\cite{Stecker:1998ib}
\bibitem{Stecker:1998ib}
  F.~W.~Stecker and M.~H.~Salamon,
  % ``Photodisintegration of ultrahigh energy cosmic rays: A new
  %determination,''
  Astrophys.\ J.\  {\bf 512}, 521 (1999).
  %[astro-ph/9808110]. 
  %%CITATION = ASTRO-PH 9808110;%% 
  %ref 38

\bibitem{zburst_problem}
O.~E.~Kalashev, V.~A.~Kuzmin, D.~V.~Semikoz and G.~Sigl,
%``Ultra-high energy cosmic rays from neutrino emitting acceleration
%sources?,''
Phys.\ Rev.\ D  65 (2002) 103003.
%[hep-ph/0112351].
%%CITATION = HEP-PH 0112351;%%
%ref 40
%ref 39


\bibitem{Neronov:2002se}
  A.~Neronov, D.~Semikoz, F.~Aharonian and O.~Kalashev,
  %``Large scale extragalactic jets powered by very high-energy gamma rays,''
  Phys.\ Rev.\ Lett.\  {\bf 89}, 051101 (2002).
 % [astro-ph/0201410].
  %%CITATION = ASTRO-PH 0201410;%%
  %ref 40


\bibitem{Kalashev:2002kx}
  O.~E.~Kalashev, V.~A.~Kuzmin, D.~V.~Semikoz and G.~Sigl,
  %``Ultra-high energy neutrino fluxes and their constraints,''
  Phys.\ Rev.\ D {\bf 66}, 063004 (2002).
 % [hep-ph/0205050].
  %%CITATION = HEP-PH 0205050;%%
  %ref 41
  


\bibitem{clark} T.~A.~Clark, L.~W.~Brown, and J.~K.~Alexander, Nature
{\bf 228}, 847 (1970).
%ref 42


\bibitem{PB}
R.~J.~Protheroe and P.~L.~Biermann,
%``A new estimate of the extragalactic radio background and implications  for
%ultra-high-energy gamma ray propagation,''
Astropart.\ Phys.\  {\bf 6}, 45 (1996)
[Erratum-ibid.\  {\bf 7}, 181 (1997)]
%[arXiv:astro-ph/9605119].
%%CITATION = ASTRO-PH 9605119;%%
%ref 43

 %\cite{Stecker:2005qs}
\bibitem{Stecker:2005qs}
F.~W.~Stecker, M.~A.~Malkan and S.~T.~Scully,
  %``Intergalactic photon spectra from the far IR to the UV Lyman limit for  0 <
  %z < 6 and the optical depth of the universe to high energy  gamma-rays,''
  Astrophys.\ J.\  {\bf 648}, 774 (2006).
  %[astro-ph/0510449].
  %%CITATION = ASTRO-PH 0510449;%%
  %ref 44

\bibitem{Dolag:2002}
K.~Dolag, M.~Bartelmann and H.~Lesch,
Astron.\ \& Astrophys. {\bf 387}, 383 (2002).
%[astro-ph/0202272].
%%CITATION = ASTRO-PH 0202272;%%
%ref 45


\bibitem{AS2}
V. S. Berezinsky, et al, {\it ``Astrophysics of Cosmic Rays.''}
(North-Holland, Amsterdam, 1990);
T.K. Gaisser, {\it ``Cosmic Rays and Particle Physics.''}
(Cambridge University Press, Cambridge, England, 1990).
%ref 46

\bibitem{AS1.5}
R.J. Protheroe, In {\it ``Topics in cosmic-ray astrophysics''},
ed. M. A. DuVernois,
Nova Science Publishing: New York, 1999, (astro-ph/9812055);
M.A. Malkov, Ap.J. {\bf 511}, L53 (1999);
K. Mannheim, R.J. Protheroe, J. P. Rachen,
Phys. Rev. {\bf D63}, 023003 (2001).
%%CITATION = ASTRO-PH 9812398;%%
%ref 47


\bibitem{peaks}
  E.~V.~Derishev, F.~A.~Aharonian, V.~V.~Kocharovsky and V.~V.~Kocharovsky,
  %``Particle acceleration through multiple conversions from charged into
  %neutral state and back,''
  Phys.\ Rev.\ D {\bf 68}, 043003 (2003).
  %  [astro-ph/0301263].
  %%CITATION = ASTRO-PH 0301263;%%
  %ref 48



\bibitem{mono}
A.~Neronov and  D.~Semikoz, 
New Astronomy Reviews {\bf 47}, 693 (2003);
A.~Neronov, P.~Tinyakov and I.~Tkachev,
  %``TeV signatures of compact UHECR accelerators,''
  J.\ Exp.\ Theor.\ Phys.\  {\bf 100}, 656 (2005)
  [Zh.\ Eksp.\ Teor.\ Fiz.\  {\bf 100}, 744 (2005)]
  [astro-ph/0402132].
  %%CITATION = ZETFA,100,744;%%
  %ref 49

\bibitem{Berezinsky:2002vt}
V.~Berezinsky, A.~Gazizov and S.~Grigorieva,
%``Signatures of AGN model for UHECR,''
astro-ph/0210095;
%%CITATION = ASTRO-PH 0210095;%%
 %``Dip in UHECR spectrum as signature of proton interaction with CMB,''
  Phys.\ Lett.\  B {\bf 612}, 147 (2005).
 % [arXiv:astro-ph/0502550].
 %ref 50

\bibitem{EGRET}
P.~Sreekumar {\it et al.}  [EGRET Collaboration],
%``EGRET observations of the extragalactic gamma ray emission,''
Astrophys.\ J.\  {\bf 494}, 523 (1998).
%[astro-ph/9709257].
%%CITATION = ASTRO-PH 9709257;%%
%ref 51

\bibitem{AGASA_clusters}
M.~Takeda {\it et al.},
%``Small-scale anisotropy of cosmic rays above 10**19-eV observed with  the
%Akeno Giant Air Shower Array,''
Astrophys.\ J.\  {\bf 522}, 225 (1999).
%[astro-ph/9902239].
%%CITATION = ASTRO-PH 9902239;%%
%ref 52

\bibitem{HiRes_clusters}
R.~U.~Abbasi {\it et al.}  [HIRES],
%``Study of small-scale anisotropy of ultrahigh energy cosmic rays observed in
%stereo by HiRes,''
Astrophys.\ J.\  {\bf 610}, L73 (2004).
[astro-ph/0404137].
%%CITATION = ASTRO-PH 0404137;%%
%ref 53

\bibitem{agasa_hires}
R.~U.~Abbasi {\it et al.} [The High Resolution Fly's Eye Collaboration
                  (HIRES)],
%``Search for point sources of ultra-high energy cosmic rays above 40-EeV using
%a maximum likelihood ratio test,''
astro-ph/0412617.
%%CITATION = ASTRO-PH 0412617;%%
%ref 54


\bibitem{agasa_hires_ok}
H.~Yoshiguchi, S.~Nagataki and K.~Sato,
  %``Statistical significance of small scale anisotropy in arrival  directions
  %of ultra-high energy cosmic rays,''
  Astrophys.\ J.\  {\bf 614}, 43 (2004);
 % [arXiv:astro-ph/0404411].
%%CITATION = ASTRO-PH 0404411;%%
M.~Kachelriess and D.~Semikoz,
  %``Ultra-high energy cosmic rays from a finite number of point sources,''
  Astropart.\ Phys.\  {\bf 23}, 486 (2005).
 % [arXiv:astro-ph/0405258].
 %ref 55


\bibitem{sources} 
E.~Waxman, K.~B.~Fisher and T.~Piran,
%``The signature of a correlation between $>10^{19}$~eV cosmic ray sources  and
%large scale structure,''
Astrophys.\ J.\  {\bf 483}, 1 (1997);
%[astro-ph/9604005].
%%CITATION = ASTRO-PH 9604005;%%
S.~L.~Dubovsky, P.~G.~Tinyakov and I.~I.~Tkachev,
%``Statistics of clustering of ultra-high energy cosmic rays and the  number of
%their sources,''
Phys.\ Rev.\ Lett.\  {\bf 85}, 1154 (2000);
%[astro-ph/0001317].
%%CITATION = ASTRO-PH 0001317;%%
Z.~Fodor and S.~D.~Katz,
%``Ultra high energy cosmic rays from compact sources,''
Phys.\ Rev.\ D {\bf 63}, 023002 (2001);
%[hep-ph/0007158].
%%CITATION = HEP-PH 0007158;%%
H.~Yoshiguchi, S.~Nagataki, S.~Tsubaki and K.~Sato,
%``Small scale clustering in isotropic arrival distribution of ultra-high
%energy cosmic rays and implications for their source candidate,''
Astrophys.\ J.\  {\bf 586}, 1211 (2003)
[Erratum-ibid.\  {\bf 601}, 592 (2004)];
%[astro-ph/0210132].
%%CITATION = ASTRO-PH 0210132;%%
%
H.~Yoshiguchi, S.~Nagataki and K.~Sato,
%``Arrival distribution of ultra-high energy cosmic rays: Prospects for  the
%future,''
Astrophys.\ J.\  {\bf 592}, 311 (2003);
%[astro-ph/0302508].
%%CITATION = ASTRO-PH 0302508;%%
P.~Blasi and D.~De Marco,
%``The small scale anisotropies, the spectrum and the sources of ultra  high
%energy cosmic rays,''
Astropart.\ Phys.\  {\bf 20}, 559 (2004).
%[astro-ph/0307067].
%%CITATION = ASTRO-PH 0307067;%%
%ref 56

\bibitem{Fodor-K-R}  Z.~Fodor, S.~D.~Katz and A.~Ringwald,
  %``Determination of absolute neutrino masses from Z-bursts,''
  Phys.\ Rev.\ Lett.\  {\bf 88}, 171101 (2002)
  %[hep-ph/0105064] 
  and 
  %``Relic neutrino masses and the highest energy cosmic rays,''
  JHEP {\bf 0206}, 046 (2002).
%  [hep-ph/0203198].
%ref 57

\bibitem{HiRes-table} http://www.physics.rutgers.edu/\%7Edbergman/HiRes-Monocular-Spectra-200702.html
%ref 58
 
\bibitem{Poisson-errors}
 G. Cowan ``Statistical data analysis",  Oxford University Press, 1998, Section 9.4. 
 %ref 59

\bibitem{statistics} 
S. Baker and R.D. Cousins, Nucl. Instrum. Methods {\bf221}, 437 (1984); Particle Data Group's Statistics Review (2004).
%ref 60

\bibitem{Albuquerque:2005nm}
  I.~F.~M.~Albuquerque and G.~F.~Smoot,
  %``GZK cutoff distortion due to the energy error distribution shape,''
  Astropart.\ Phys.\  {\bf 25}, 375 (2006).
 % [astro-ph/0504088].
  %%CITATION = APHYE,25,375;%%
  %ref 61

\bibitem{1.2factorAGASA}
  P.~Homola {\it et al.},
  %``Simulation of ultra-high energy photon showers with PRESHOWER,''
  Nucl.\ Phys.\ Proc.\ Suppl.\  {\bf 151} (2006) 119;
  %%CITATION = NUPHZ,151,119;%%
  M.~Risse {\it et al.},
  %``Upper limit on the photon fraction in highest-energy cosmic rays from
  %AGASA data,''
  Phys.\ Rev.\ Lett.\  {\bf 95}, 171102 (2005).
 % [arXiv:astro-ph/0502418].
  %%CITATION = PRLTA,95,171102;%%

\bibitem{AgasaYakutskLimit}  
  G.~I.~Rubtsov {\it et al.},
  % ``Upper limit on the ultra-high-energy photon flux from AGASA and Yakutsk
  %data,''
  Phys.\ Rev.\ D {\bf 73}, 063009 (2006).
  %[astro-ph/0601449].
  %%CITATION = ASTRO-PH 0601449;%%
  
\bibitem{Yakutsk}A.~V.~Glushkov {\it et al.}
%D.~S.~Gorbunov, I.~T.~Makarov, M.~I.~Pravdin, G.~I.~Rubtsov, I.~E.~Sleptsov and S.~V.~Troitsky,
  %``Constraining the fraction of primary gamma rays at ultra-high energies from
  %the muon data of the Yakutsk extensive-air-shower array,''
  JETP Lett.\  {\bf 85}, 131 (2007).
%  [arXiv:astro-ph/0701245].
  %%CITATION = JTPLA,85,131;%%
 
\bibitem{sms05}
F.~W.~Stecker, M.~A.~Malkan and S.~T.~Scully,
  %``Intergalactic photon spectra from the far IR to the UV Lyman limit for  0 <
  %z < 6 and the optical depth of the universe to high energy  gamma-rays,''
  Astrophys.\ J.\  {\bf 648}, 774 (2006).
 % [arXiv:astro-ph/0510449].
 
 \bibitem{dp90}
J.~S.~Dunlop and J.~A.~Peacock, MNRAS {\bf 247}, 19 (1990). 
 
 \bibitem{Inoue:2007kn}
  S.~Inoue, G.~Sigl, F.~Miniati and E.~Armengaud,
  %``Ultrahigh energy cosmic rays as heavy nuclei from cluster accretion
  %shocks,''
  astro-ph/0701167.
 
\bibitem{ANITA-Light} 
  S.~W.~Barwick {\it et al.}  [ANITA Collaboration],
  %``Constraints on cosmic neutrino fluxes from the ANITA experiment,''
  Phys.\ Rev.\ Lett.\  {\bf 96}, 171101 (2006).
 % [arXiv:astro-ph/0512265].
  %%CITATION = PRLTA,96,171101;%%
  
  \bibitem{AMANDA}
  M.~Ackermann {\it et al.}  [AMANDA Collaboration],
  %``Search for neutrino induced cascades with AMANDA,''
  Astropart.\ Phys.\  {\bf 22}, 127 (2004).
  %[arXiv:astro-ph/0405218].
  %%CITATION = APHYE,22,127;%%
  
  \bibitem{RICE}
  A.~Bean  [RICE Collaboration],
 %``RICE limits on the diffuse ultra-high energy neutrino flux,''
 AIP Conf.\ Proc.\  {\bf 870} (2006) 212.

\end{thebibliography}
\end{document}